\documentclass[letterpaper, 10 pt, conference]{ieeeconf}

\renewcommand{\baselinestretch}{0.95}
\usepackage{wrapfig}

\usepackage{savesym}
 
\usepackage{amsmath,amssymb,amsfonts,amsthm}
\usepackage{comment}
\usepackage{xspace}
\usepackage{scalerel}
\usepackage{xcolor}
\usepackage[font=footnotesize]{caption}
\usepackage[font=footnotesize]{subcaption}
\usepackage{multicol}
\usepackage{bm}
\usepackage{enumerate}
\usepackage{booktabs} 
\usepackage{graphicx}
\usepackage{bbm}
\usepackage{dsfont}
\usepackage{listing}
\usepackage[ruled, vlined,linesnumbered, noend]{algorithm2e}
\usepackage[english]{babel}
\usepackage{csquotes}
\usepackage{lipsum}

\makeatletter
\let\NAT@parse\undefined
\makeatother
\usepackage[colorlinks=true, bookmarks=true, pagebackref=true,citecolor=blue]{hyperref}

\IEEEoverridecommandlockouts
\def\BibTeX{{\rm B\kern-.05em{\sc i\kern-.025em b}\kern-.08em
				T\kern-.1667em\lower.7ex\hbox{E}\kern-.125emX}}

\graphicspath{ {fig/} }
\newif \ifarXiv
\arXivtrue

\begin{document}

\title{\Large \bf Sample Complexity of Probabilistic Roadmaps via $\bm{\epsilon}$-nets }

\author{Matthew Tsao, Kiril Solovey, and Marco Pavone
\thanks{The authors are with the Autonomous Systems Laboratory, Stanford University, CA~94305, USA. This work was supported in part by NSF, Award Number: 1931815.}
}

\maketitle

\newtheorem{theorem}{Theorem}
\newtheorem{corollary}{Corollary}
\newtheorem{lemma}{Lemma}
\newtheorem{observation}{Observation}
\newtheorem{proposition}{Proposition}
\newtheorem{claim}{Claim}
\newtheorem{fact}{Fact}
\newtheorem{assumption}{Assumption}
\theoremstyle{definition}
\newtheorem{definition}{Definition}
\newtheorem{remark}{Remark}
\newtheorem{example}{Example}

\newcommand{\abs}[1]{\left| #1 \right|}
\newcommand{\bigpar}[1]{\left( #1 \right)}
\newcommand{\bigbra}[1]{\left[ #1 \right]}
\newcommand{\bigbrace}[1]{\left\{ #1 \right\}}
\newcommand{\braket}[1]{\left\langle #1 \right\rangle}
\newcommand{\bigceil}[1]{\left\lceil #1 \right\rceil}
\newcommand{\bigfloor}[1]{\left\lfloor #1 \right\rfloor}
\newcommand{\casewise}[1]{\left\{ #1 \right.}
\newcommand{\norm}[1]{\left| \left| #1 \right| \right|}
\newcommand{\iid}{\overset{\text{i.i.d}}{\sim}}
\newcommand{\highlight}[2]{\color{#1} #2 \color{black}}

\newcommand{\arXivver}{~\cite{TsaoSoloveyPavone20a} }

\newcommand{\kiril}[1]{{\color{blue}({\bf Kiril:} #1)}}
\newcommand{\todo}[1]{{\color{red}{\bf TODO:} #1}}
\newcommand{\matt}[1]{{\color{magenta}{\bf Matt:} #1}}

\newcommand{\cpp}{C\raise.08ex\hbox{\tt ++}\xspace}

\def\P{\mathcal{P}} \def\C{\mathcal{C}} \def\H{\mathcal{H}}
\def\F{\mathcal{F}} \def\U{\mathcal{U}} \def\L{\mathcal{L}}
\def\O{\mathcal{O}} \def\I{\mathcal{I}} \def\S{\mathcal{S}}
\def\G{\mathcal{G}} \def\Q{\mathcal{Q}} \def\I{\mathcal{I}}
\def\T{\mathcal{T}} \def\L{\mathcal{L}} \def\N{\mathcal{N}}
\def\V{\mathcal{V}} \def\B{\mathcal{B}} \def\D{\mathcal{D}}
\def\W{\mathcal{W}} \def\R{\mathcal{R}} \def\M{\mathcal{M}}
\def\X{\mathcal{X}} \def\A{\mathcal{A}} \def\Y{\mathcal{Y}}
\def\L{\mathcal{L}} \def\K{\mathcal{K}}

\def\dS{\mathbb{S}} \def\dT{\mathbb{T}} \def\dC{\mathbb{C}}
\def\dG{\mathbb{G}} \def\dD{\mathbb{D}} \def\dV{\mathbb{V}}
\def\dH{\mathbb{H}} \def\dN{\mathbb{N}} 
\def\dR{\mathbb{R}} \def\dM{\mathbb{M}} \def\dm{\mathbb{m}}
\def\dB{\mathbb{B}} \def\dI{\mathbb{I}} \def\dM{\mathbb{M}}
\def\dZ{\mathbb{Z}} \def\dP{\mathbb{P}}

\def\len{\textup{len}}
\def\vol{\textup{vol}}

\def\eps{\varepsilon}

\def\limn{\lim_{n\rightarrow \infty}}

\def\indicator{\mathds{1}}
\def\eqq{\coloneqq}

\def\Reals{\mathbb{R}}
\def\Naturals{\mathbb{N}}
\renewcommand{\leq}{\leqslant}
\renewcommand{\geq}{\geqslant}
\newcommand{\compl}{\mathrm{Compl}}

\newcommand{\np}{{\sc np}\xspace}

\def\dt{\,\mathrm{d}t}
\def\dx{\,\mathrm{d}x}
\def\dy{\,\mathrm{d}y}
\def\ds{\,\mathrm{d}s}
\def\drho{\,\mathrm{d}\rho}

\def\x{\bm{x}}
\def\y{\bm{y}}
\def\z{\bm{z}}
\def\k{\bm{\kappa}}

\def\barE{\bar{E}}

\newcommand{\prm}{{\tt PRM}\xspace}
\newcommand{\prmstar}{{\tt PRM}$^*$\xspace}
\newcommand{\rrt}{{\tt RRT}\xspace}
\newcommand{\est}{{\tt EST}\xspace}
\newcommand{\rrtstar}{{\tt RRT}$^*$\xspace}
\newcommand{\rrg}{{\tt RRG}\xspace}
\newcommand{\btt}{{\tt BTT}\xspace}
\newcommand{\bit}{{\tt BIT}$^*$\xspace}

\newcommand{\lbtrrt}{{\tt LBT-RRT}\xspace}
\newcommand{\mplb}{{\tt MPLB}\xspace}
\newcommand{\spars}{{\tt SPARS2}\xspace}
\newcommand{\rsec}{{\tt RSEC}\xspace}
\newcommand{\bfmt}{{\tt BFMT}$^*$\xspace}
\newcommand{\astar}{{\tt A}$^*$\xspace}

\newcommand{\fmt}{{\tt FMT}$^*$\xspace}
\newcommand{\mstar}{{\tt M}$^*$\xspace}
\newcommand{\drrt}{{\tt dRRT}\xspace}
\newcommand{\drrtstar}{{\tt dRRT}$^*$\xspace}

\newcommand{\xstart}{x_{\textup{start}}}
\newcommand{\xgoal}{x_{\textup{goal}}}

\let\oldnl\nl
\newcommand{\nonl}{\renewcommand{\nl}{\let\nl\oldnl}}


\begin{abstract}
We study fundamental theoretical aspects of probabilistic roadmaps (PRM) in the finite time (non-asymptotic) regime. In particular, we investigate how completeness and optimality guarantees of the approach are influenced by the underlying deterministic sampling distribution $\bm{\X}$ and  connection radius $\bm{r>0}$.  We develop the notion of $\bm{(\delta,\epsilon)}$-completeness of the parameters $\bm{\X, r}$, which indicates that for every motion-planning problem of clearance at least $\bm{\delta>0}$, PRM using $\bm{\X, r}$ returns a solution no longer than $\bm{1+\epsilon}$ times the shortest $\bm{\delta}$-clear path. 
Leveraging the concept of $\bm{\epsilon}$-nets, we characterize in terms of lower and upper bounds the number of samples needed to guarantee $\bm{(\delta,\epsilon)}$-completeness. This is in contrast with previous work which mostly considered the asymptotic regime in which the number of samples tends to infinity. In practice, we propose a sampling distribution inspired by $\bm{\epsilon}$-nets that achieves nearly the same coverage as grids while using fewer samples. 
\end{abstract}


\section{Introduction}\label{sec:intro}
The Probabilistic Roadmap Method (\prm)~\cite{KavrakiETAL96} is one of the most widely used sampling-based technique for motion planning. \prm  generates a graph approximation of the full free space of the problem, by generating a set of configuration samples and connecting nearby samples when it is possible to move between configurations without collision using straight-line paths. \prm is particularly suitable in multi-query settings, where the workspace environment needs to be preprocessed to answer multiple queries consisting of different start and goal points. Recently \prm has been applied to challenging robotic settings, including manipulation planning~\cite{KimmelETAL18}, inspection planning and coverage~\cite{FuETAL19}, task planning~\cite{GarrettETAL18b}, and multi-robot motion planning~\cite{ShomeETAL19, HonigETAL18}. \prm is instrumental in many modern single-query planners, which implicitly maintain a \prm graph 
and return a solution that minimizes the path's cost~\cite{JSCP15,StaETAL15,SH15,GSB15,SolHal16}. 

Extensive study of \prm 's theoretical properties quickly followed its inception. The first question that occupied the research community was whether \prm guarantees to find a solution if one exists~\cite{KavrakiETAL98,LadKav04b,ChaudhuriKoltun09}, and later on the quality of the returned solution. Several works have established the magnitude of the connection radius $r$ sufficient to guarantee the convergence of the solution returned by \prm to an optimal solution~\cite{JSCP15,KF11,SK19}. However, the majority of works addressing those two questions consider in their analysis the (somewhat unrealistic) asymptotic regime, where the number of samples  $n$ tends to infinity. \emph{The question of what are the smallest values of $n, r$ to guarantee a high-quality solution in practice, i.e., when $n$ is fixed, remains open.} \vspace{5pt}

\noindent \textbf{Statement of Contributions:} In this work we make progress toward addressing the aforementioned question. In particular, we study how the sample set $\X$, and its cardinality $n$, as well as the size of the connection radius $r$, affect completeness and optimality guarantees of \prm. We develop the notion of  ${(\delta,\epsilon)}$-completeness of the parameters ${\X, r}$, which indicates that for every motion-planning problem of clearance  ${\delta>0}$, \prm using $\X, r$ returns a solution no longer than ${1+\epsilon}$ times the shortest path with at least $\delta$ clearance from the obstacles. 

The concept of $\epsilon$-nets~\cite{MustafaVaradarajan16} plays a key role in our contributions in both the theory and application. From a theoretical perspective, we leverage properties of $\epsilon$-nets to characterize in terms of lower and upper bounds the sample size and connection radius needed to guarantee ${(\delta,\epsilon)}$-completeness. This is in contrast with previous work which mostly considered the asymptotic regime in which the number of samples tends to infinity. 
From an application perspective, we leverage properties of $\epsilon$-nets via a template method to produce sample sets that efficiently cover the workspace. We observe empirically that these sample sets offer nearly the same coverage as grids while using fewer samples. Grids are an important baseline because they are used widely in practice and offer better coverage (i.e., dispersion) than uniform random sampling~\cite{LucETAL17}. The increased efficiency provided by the template method over grids can improve the runtime of \prm and related algorithms. 

This paper is organized as follows. Related work is surveyed in Section \ref{sec:related}. Preliminaries are discussed in Section~\ref{sec:preliminaries}. Section~\ref{sec:theorems} presents the main theoretical contributions and Section~\ref{sec:algorithms} presents proof sketches for some of the results. Numerical experiments comparing the efficiency of our $\epsilon$-net based template method to grids are presented in Section~\ref{sec:experiments}. We summarize our work and discuss future directions in Section~\ref{sec:conclusion}. 

\section{Related work}\label{sec:related}
We provide a literature review of results concerning the theoretical properties of \prm. The majority of results apply to the setting of a Euclidean configuration spaces, and samples that are generated in a uniform and random fashion.


The the study of asymptotic optimality  in sampling-based planning was initiated in~\cite{KF11}. 
This paper proves that if $r>\gamma\left(\tfrac{\log n}{n}\right)^{1/d}$, for some constant $\gamma>0$, then the length of the solution returned by \prm converges asymptotically almost surely (a.a.s.), as the number of samples $n\rightarrow \infty$, to length cost of the robust optimal solution. Such a connection radius leads to a graph of size $\Theta(n\log n)$, in contrast to a size of $\Theta(n^2)$ induced by a constant radius (as in~\cite{KavrakiETAL96}). Subsequent work managed to further reduce the constant $\gamma$~\cite{JSCP15, SoloveyETAL18}. A recent paper~\cite{SK19} establishes the existence of a critical connection radius $r^*=\gamma'\left(\frac{1}{n}\right)^{1/d}$, where $\gamma'$ is constant: if $r<r^*$ then \prm is guaranteed to fail (even when $n\rightarrow \infty$), and if $r>r^*$ then it is guaranteed to converge a.a.s. to a near-optimal solution. A finite-time analysis of \prm, providing probabilistic bounds for achieving a given stretch factor $1+\epsilon$ for a fixed number of samples $n$ and a specific connection radius of the aforementioned form was established in~\cite{DobsonBekris13,DobETAL15}. An algorithm has stretch factor $\beta$ if it produces a solution whose length is no more than $\beta$ times the length of the optimal solution. 

The aforementioned results assume a uniform random sampling scheme. A recent work~\cite{LucETAL17} establishes that using a low-dispersion deterministic sampling scheme (e.g., Halton and Sukharev sequences), asymptotic optimality is achieved with a radius as small as $f(n)\left(\frac{1}{n}\right)^{1/d}$, for any  
$\lim_{n\rightarrow \infty}f(n)=\infty$. 

Two recent works~\cite{SchETAL15, SchETAL15b} consider non-Euclidean systems and develop sufficient conditions for asymptotic optimality with uniform random sampling.

\section{Preliminaries}\label{sec:preliminaries}
We provide several basic definitions. Given two points $x,y\in \dR^d$, denote by $\norm{x-y}_p := (\sum_{i=1}^d \abs{x_i - y_i}^p)^{1/p}$ the $\ell_p$ distance between them. When $p=2$ we obtain the standard Euclidean distance. We denote the $d$-dimensional $\ell_p$ ball with radius $r>0$ centered at $x\in \dR^d$ as $B_p(x,r) := \{y : \|x-y\|_p \leq r\}$. For a Euclidean set $\X$, $\text{co} \bigpar{ \X }$  denotes its convex hull and $\text{vol}(\X)$ to denote its volume. 
	
\subsection{Motion planning}
Denote by $\C$ the configuration space of the robot, which, by rescaling we will assume is $[0,1]^d$ throughout this paper. The free space $\F \subset \C$ denotes all collision-free configurations. A motion-planning problem is then specified by the tuple $\bigpar{\F, x_{\text{start}},x_{\text{goal}}}$. The objective is to find a (continuous) path $p : [0,1] \rightarrow [0,1]^d$ that a) moves the robot from the start to goal location, i.e. $p(0) = x_{\text{start}}, p(1) = x_{\text{goal}}$ and b) avoids collisions with obstacles, i.e. $p(t) \in \F$ for all $t \in [0,1]$.  

We measure the quality of a path $p$ by its length $\ell(p)$. A crucial property of paths in sampling-based planning is the notion of clearance.  A motion-planning problem $\bigpar{\F, x_{\text{start}},x_{\text{goal}}}$ has $\delta$-clearance if there exists a path $p : [0,1] \rightarrow [0,1]^d$ with $p(0) = x_{\text{start}}$, $p(1) = x_{\text{goal}}$, and  $\bigcup_{t \in [0,1]} B_2(p(t),\delta) \subset \F$.

\subsection{Probabilistic Roadmaps}
We provide a formal definition of the Probabilistic Roadmaps Method (\prm). For a given motion-planning problem $\M:=(\F,\xstart,\xgoal)$, \prm generates a graph $G_{\M}$, whose vertices and edges represent collision-free configurations and straight-line paths connecting configurations, respectively. 

The \prm graph induced by $\M,\X,r$ is denoted by $G_\M(\mathcal{X},r) = (V,E)$. The vertices $V$ are all collision-free configurations in $\X \cup \bigbrace{\xstart , \xgoal}$. 
The (undirected) edges $E$ connect between every pair of vertices $u,v\in V$ such that (i) the distance between them is at most $r$, and (ii) the straight-line path between them is entirely collision free. Formally,

\vspace{-10pt} 
{\small\begin{align*}
V &:= \left(\X \cup \bigbrace{\xstart , \xgoal}\right)\cap \F, \\
E &:= \bigbrace{ (u,v) \in V \times V : \norm{u-v}_2 \leq r, \textup{co} \bigpar{\bigbrace{u,v}} \subset \F }.
\end{align*}}

\vspace{-15pt}
\subsection{Sampling distributions}
The choice of sample set $\X$ has significant implications on the properties of $G_{\M}(\X,r)$. We discuss two types of sample sets, namely $\epsilon$-nets and grids, that are particularly useful for \prm. We use \textit{sample set} and \textit{sample distribution} interchangeably when referring to $\X$. 

\begin{definition}[$\epsilon$-nets]
A set $\B \subset \mathbb{R}^d$ is a $\epsilon$-net for a set $\A \subset \mathbb{R}^d$ if for every $a \in \A$, there exists a $b \in \B$ so that $\norm{a-b}_2 \leq \epsilon$. 
\end{definition}

As far as motion planning is concerned, $\epsilon$-nets are good candidates for deterministic sampling schemes since by definition they have uniformly dense coverage over the entire space. The low dispersion sequences mentioned in~\cite{LucETAL17} (i.e. Sukharev Grids) are special cases of $\epsilon$-nets, and we will show that studying these objects in full generality leads to improved covering efficiency. \textit{Minimal} $\epsilon$-nets, i.e. nets with the smallest possible cardinality, are then good candidates for \textit{efficiently} covering the entire space. We will make these intuitions formal in the coming sections. Grids have long seen use as sample sets in motion planning~\cite{lavalle06}, so we use grids as benchmarks against which we compare the performance of sample sets inspired by $\epsilon$-nets. We define a Sukharev grid of $[0,1]^d$ with spacing $w$ to be the following collection of points:

\vspace{-10pt}
\begin{small}
\begin{align*}
\X_{\text{grid}}(w) := \bigbrace{x \in \mathbb{R}^d : \frac{x_i}{w} + \frac{1}{2} \in \bigbrace{1,2,..., \frac{1}{w}} \forall i}.
\end{align*}
\end{small}
\vspace{-10pt}

See Section~\ref{sec:theorems:gridver} for a detailed comparison between grid sampling and $\epsilon$-net sampling. 

\subsection{Completeness as a Benchmark}
We will use the following benchmark to measure the quality of samples set $\X$ and a connection radius.

\begin{definition}[$(\delta,\epsilon)$-completeness]
Given a samples set $\X$ and a connection radius $r$, we say that the pair $(\X,r)$ is  $(\delta, \epsilon)$-complete if for every $\delta$-clear $\M$ it holds that 
\[d_{G_\M(\X,r)} \bigpar{x_{\text{start}}, x_{\text{goal}}} < (1+\epsilon)\cdot 
\textup{OPT}_\delta,\]
where $d_{G_\M(\X,r)} \bigpar{x_{\text{start}}, x_{\text{goal}}}$ denotes the length of the shortest path from start to goal in the graph $G_\M(\X,r)$, and $\textup{OPT}_\delta$ is the length of the shortest $\delta$-clear solution to $\M$. 
\end{definition}

Note that $\epsilon = \infty$ corresponds to the case where finding a feasible 
path is the only objective. 

\section{Sample Complexity of PRM}\label{sec:theorems}
Our main objective is to study the sample complexity of \prm. For a clearance level $\delta>0$ and a stretch tolerance $\epsilon>0$, we wish to find comprehensive conditions on when a sample set $\X$ and connection radius $r$ are $(\delta,\epsilon)$-complete.  
The properties of $\epsilon$-nets are central to our contributions to this goal, as well as our experimental results in Section~\ref{sec:experiments}.

We focus on deterministic sampling but mention that our techniques can be used to derive results for i.i.d.\ sampling procedures as well, which show a poorer performance of i.i.d.\ sampling, in line with the findings of~\cite{LucETAL17}. As required by our main objective, all of our results are presented in the finite-sample, non-asymptotic setting. 



We will prove sufficient conditions on sample complexity in a constructive way by exhibiting algorithms that achieve the conditions. To this end, the \hyperref[alg:nmp]{\texttt{Epsilon Net Sampling}} (\texttt{ENS}) procedure presented in Algorithm~\ref{alg:nmp} will be used to certify the sufficient conditions. \hyperref[alg:nmp]{\texttt{ENS}} leverages properties of $\epsilon$-nets to determine a connection radius $r$ and net resolution $\delta_{\text{min}}$. It then calls as a subroutine the \hyperref[alg:epsnet]{\texttt{Build-Net}} algorithm, which constructs a $\delta_{\text{min}}$-net of the configuration space by including points until a $\delta_{\text{min}}$-net is obtained, while ensuring that all points included are at pairwise distance at least $\delta_{\text{min}}$. In the next section, we will prove the efficacy of \hyperref[alg:nmp]{\texttt{ENS}} which in turn gives sufficient conditions on the complexity of \prm.  

\begin{algorithm}[t]\SetAlgoSkip{}
\textbf{Input}: Set $\A$, cover radius $\epsilon > 0$\;
\textbf{Output}: An $\epsilon$-net $\B$ for $\A$\;
$\mathcal{B} \leftarrow \emptyset$\;
\While{$\A \setminus \cup_{b \in \B} B_2(y,\epsilon) \neq \emptyset$}{
Find $y \in \A \setminus \cup_{b \in \B} B_2(b,\epsilon)$\;
$\B \leftarrow \B \cup \bigbrace{y}$\;
}
\Return $\B$. \;
\caption{\texttt{Build-Net}}\label{alg:epsnet}
\end{algorithm}

\begin{algorithm}[t]\SetAlgoSkip{}
\textbf{Input}: Sample size $n$, workspace dimension $d$, stretch tolerance $\epsilon > 0$, $\delta$ clearance parameter\;
\textbf{Output}: Sample set $\X$ and connection radius $r$\;
$\alpha \gets \tfrac{\epsilon}{\sqrt{1+\epsilon^2}}$\;
$n_\delta \gets\min {\small\left\{ \sqrt{\pi d} \bigpar{ \sqrt{\tfrac{2d}{\pi e}} \cdot \tfrac{ 1-(2-\alpha)\delta}{ \alpha \delta}}^d, n\right\}}$\;
$\delta_{\text{min}}\! \gets\! {\small\sqrt{\tfrac{2d}{\pi e}} \!\bigpar{ \tfrac{\sqrt{\pi d}}{n_\delta} }^{\tfrac{1}{d}}\! \bigbra{\! \alpha\! -\! (2-\alpha)\! \sqrt{\tfrac{2d}{\pi e}} \bigpar{ \tfrac{\sqrt{\pi d}}{n_\delta} }^{\tfrac{1}{d}} }^{-1}}$\;
$r \gets 2 \bigpar{\alpha + \sqrt{1-\alpha^2}} \delta_{\min}$\;
$\X \gets \hyperref[alg:epsnet]{\texttt{Build-Net}}([\delta_{\min},1-\delta_{\min}]^d, \delta_{\min})$\;
\Return $(\X,r)$. 
\caption{\texttt{Epsilon Net Sampling} (\texttt{ENS})}\label{alg:nmp}
\end{algorithm}

\subsection{Main Results}\label{sec:deterministic}
The following theorem provides a necessary condition on the size of a sample set $\X$ and radius $r$ to be $(\delta,\epsilon)$-complete.

\begin{theorem}[Necessary Conditions]\label{thm:detness}
Let $\X \subset [0,1]^d$ be a set of $n$ points and let $\delta>0$. If $(\X,r)$ is $(\delta,\infty)$-complete then 
{\small\begin{align*}
n &\geq \sqrt{\frac{e}{2}} \bigpar{1 - \frac{2\delta}{1-2\delta}}^2 \bigpar{ \sqrt{\frac{d-1}{2\pi e}} \cdot \frac{\bigpar{1-2\delta}}{\delta}}^d \\
\text{ and } r &\geq (1-2\delta) \bigpar{\sqrt{\pi d}}^{1/d} \sqrt{\frac{d}{2 \pi e }} \bigpar{ \frac{1}{n} }^{1/d}.
\end{align*}}
\end{theorem}

In particular, any sample set smaller than the lower bound, regardless of $r$ and how the points are chosen, cannot be $(\delta,\infty)$-complete. Next, we focus our efforts in finding $(\X,r)$ with size and radius comparable to the lower bound from Theorem~\ref{thm:detness}, for which $(\delta,\epsilon)$-completeness is guaranteed. The following theorem leverages \hyperref[alg:nmp]{\texttt{ENS}} to achieve this goal. 

\begin{theorem}[Sufficient Conditions]\label{thm:detsuff} 
  Let $\epsilon > 0,\delta >0$. If
  {\small\begin{align*}
    n &\geq \sqrt{\pi d} \bigpar{ \sqrt{\frac{2d}{\pi e}} \cdot \frac{ 1-(2-\alpha)\delta}{ \alpha \delta}}^d \text{ and } \\
    r &\geq 2 \bigpar{ 1 + \frac{1}{\epsilon}} \bigpar{\sqrt{\pi d}}^{1/d} \sqrt{\frac{d}{2 \pi e }} \bigpar{ \frac{1}{n} }^{1/d}
  \end{align*}}where $\alpha = \frac{\epsilon}{\sqrt{1+\epsilon^2}}$, then 
$\texttt{ENS}(n,d,\delta)$ is $(\delta,\epsilon)$-complete. 
\end{theorem}


\ifarXiv
See Sections~\ref{pf:thm:detness} and~\ref{pf:thm:detsuff} for proofs of Theorems~\ref{thm:detness} and~\ref{thm:detsuff}. 
\fi

\subsection{Discussion}
A few comments are in order. The connection radius condition in Theorem~\ref{thm:detsuff} generalizes the works of~\cite{KF11} and~\cite{LucETAL17} to finite sample settings. Concretely, choosing $\epsilon = (\log n)^{-1/d}$ achieves asymptotic optimality with a radius $r = \gamma \bigpar{ \frac{\log n}{n} }^{1/d}$, recovering the result of~\cite{KF11}. This is because $\epsilon \rightarrow 0$ as $n \rightarrow \infty$, and the minimum $\delta$ satisfying the sample size condition in Theorem~\ref{thm:detsuff} goes to zero as $n \rightarrow \infty$. More generally, for any diverging function $f(n)$, choosing $\epsilon = \frac{1}{f(n)}$ recovers the asymptotic optimality condition $r = \gamma f(n) n^{-1/d}$ from~\cite{LucETAL17}. 

Theorem~\ref{thm:detsuff} provides several other additions to the motion planning literature. It provides guarantees on the stretch factor achievable with finite $n$, which are not specified in the aforementioned works. This result is actionable in the sense that for a given $\delta$ and $\epsilon$, it provides a sample size and distribution guaranteed to find a solution with the desired quality. Conversely, Theorem~\ref{thm:detsuff} can be used to certify of hardness (or as a non-existence proof~\cite{BaschETAL01,McCarthyETAL12}): for any $n,\delta >0$ satisfying the condition in Theorem~\ref{thm:detsuff}, if there does not exist a feasible path in $\G_\M(\X,r)$ where $(\X,r) = \texttt{ENS}(n,d,\delta)$, then  $\M$ has clearance strictly less than $\delta$.

We discuss the implications of our results on the sample complexity of \prm. For small dimensions, i.e., $d \leq 4$, Theorem~\ref{thm:detsuff} shows that $O(\delta^{-d})$ samples are sufficient for $(\delta,\infty)$-completeness, since $\epsilon = \infty$ corresponds to $\alpha = 1$. Conversely, Theorem~\ref{thm:detness} shows that this sample complexity is essentially optimal (up to a multiplicative constant factor) in the sense that every $(\delta,\infty)$-complete $(\X,r)$ must have $\abs{\X} = \Omega(\delta^{-d})$. In high dimensions,  the ratio between the sufficient and necessary conditions on sample size is $\Omega(2^d \sqrt{d})$. Thus when $d$ is no longer a small constant, there is a significant gap between the achievability result of Theorem~\ref{thm:detsuff} to the lower bound on sample complexity given by Theorem~\ref{thm:detness}. \vspace{5pt}

\noindent \textbf{Evaluating the Bounds:} To give a sense of the sample size conditions specified by Theorems~\ref{thm:detness} and~\ref{thm:detsuff}, Table~\ref{tbl:sampcomp_examples} evaluates bounds for $n=|\X|$ with respect various values of $\delta,d,\epsilon$. 

From this we observe a poor scaling with clearance level $\delta$, implying that \prm with classical sampling methods is likely not the right tool for high dimensional, low clearance motion-planning problems with low error tolerance (i.e., where success must be guaranteed). For example, if $\M$ is a maze in $d=5$ dimensions with path width $0.01$ (i.e. clearance $\delta = 0.005$), then by Theorem~\ref{thm:detness}, at least $9.2 \cdot 10^{9}$ samples are needed to ensure that \prm will find a solution to the maze. A graph this size is beyond what can be stored by modern computers. 

The story is more optimistic for larger values of $\delta$ (i.e., problems with higher clearance) and lower dimensions. When $\delta = 0.25$, a feasible solution can be guaranteed (see $\epsilon = \infty$) with 8000 samples. For $d=4,\delta = 0.1$, a feasible solution can be guaranteed with $20,000$ samples. In fact, a stretch factor of $2$ (see $\epsilon = 1$) can be guaranteed in $\delta = 0.25$ clearance with a similar sample size on the order of $10^4$. Theorem~\ref{thm:detsuff} reveals that, with the right sampling scheme, \prm is guaranteed to find a solution efficiently and can do so in real time. For smaller clearance levels like $\delta = 0.1$, a solution can be guaranteed with somewhere between $10^5$ and $10^6$ samples, depending on the dimension. A solution with stretch factor $2$ can be guaranteed by inflating the sample size by an additional factor of $10$. This is no longer acceptable for real-time applications, but is still manageable for offline preprocessing if the environment will be queried many times, allowing for a cheap amortized cost. 

While many practical robotic systems are high dimensional, i.e., $d\geq 6$, some application instances naturally admit decoupling of the degrees of freedom of the system, which induces lower-dimensional configuration subspaces. For instance, manipulation problems (see, e.g., \cite{KimmelETAL18})  can be typically decomposed into a sequence of tasks where the manipulator is driving toward an object (while fixing its arms), then moves an arm towards the object, and finally grasps it by actuating its fingers. Additionally, in some settings prior knowledge about the structure of the environment or a lower-dimensional space can inform sampling in the full configuration space, which can lead to more informative sampling distributions~\cite{ReidETAL16, IchterHarrisonEtAl2018, IchterPavone19, ChoudhuryBAKRSD18}. Therefore, while \prm with classical sampling methods may not be the right tool for solving an entire motion planning problem, it can still be useful for solving subtasks in heirarchical task models or covering latent spaces efficiently. 


\begin{table}
\caption{Sample Complexity Examples}\label{tbl:sampcomp_examples}
\centerline{
\begin{tabular}{|cc|c|ccc|}
\hline
$\delta$ & $d$ & Thm.~\ref{thm:detness} LB & & Thm.~\ref{thm:detsuff} UB & \\
 &  & & $\epsilon = \infty$ & $\epsilon = 1$ & $\epsilon = 0.25$\\
\hline
$0.25$ & $4$ & $0$ & $252$ & $669$ & $22737$\\
\hline
$0.25$ & $5$ & $0$ & $1430$ & $4837$ & $3.9 \cdot 10^5$\\
\hline
$0.25$ & $6$ & $0$ & $8781$ & $37930$ & $7.5 \cdot 10^6$\\
\hline
$0.1$ & $4$ & $82$ & $20411$ & $7.15 \cdot 10^4$ & $4.2 \cdot 10^{6}$\\
\hline
$0.1$ & $5$ & $570$ & $3.48 \cdot 10^{5}$ & $1.66 \cdot 10^6$ & $2.6 \cdot 10^{8}$ \\
\hline
$0.1$ & $6$ & $4313$ & $6.41 \cdot 10^{6}$ & $4.19 \cdot 10^7$ & $1.8 \cdot 10^{10}$\\
\hline
$0.05$ & $4$ & $2983$ & $4.1 \cdot 10^{5}$ & $1.52 \cdot 10^6$ & $9.9 \cdot 10^{7}$\\
\hline
$0.05$ & $5$ & $46201$ & $1.46 \cdot 10^{7}$ & $7.62 \cdot 10^7$ & $1.4 \cdot 10^{10}$\\
\hline
$0.05$ & $6$ & $7.86\cdot 10^5$ & $5.67 \cdot 10^{8}$ & $4.13 \cdot 10^9$ & $2.2 \cdot 10^{12}$\\
\hline
\end{tabular}
}\vspace{-15pt}
\end{table}

\subsection{Implications for Grid Sampling}\label{sec:theorems:gridver}
The achievability result from Theorem~\ref{thm:detsuff} is obtained by using \hyperref[alg:nmp]{\texttt{ENS}}. 
In this section we provide sufficient conditions on sample size for \prm to be $(\delta,\epsilon)$-complete when using grid sampling. Such a result serves as a benchmark for the proposed sampling algorithm \hyperref[alg:nmp]{\texttt{ENS}}, and may be of independent interest as grids are commonly used in motion planning. 

\begin{corollary}[Sufficient Conditions for Grid Sampling]\label{cor:detsuffgrid}
  Take $r$ to satisfy the connection radius condition given in Theorem~\ref{thm:detsuff}. Then $\bigpar{ \X_{\text{grid}} \bigpar{ \frac{2 \alpha \delta}{\sqrt{d}}}, r }$ is $(\delta,\epsilon)$-complete and 
\begin{small}
\begin{align*}
\abs{ \X_{\text{grid}} \bigpar{ \frac{2 \alpha \delta}{\sqrt{d}}} } = \bigpar{ \frac{\sqrt{d}}{2} \cdot \frac{ 1-2\delta}{ \alpha \delta}}^d, \text{ where } \alpha = \frac{\epsilon}{\sqrt{1+\epsilon^2}}.
\end{align*}
\end{small}
\end{corollary}

\ifarXiv
See Section~\ref{pf:gridsampleruntime} for a proof of corollary~\ref{cor:detsuffgrid}. 
\fi

One natural comparison to make then is between the quality of coverage offered by grids versus $\epsilon$-net approaches like \hyperref[alg:epsnet]{\texttt{Build-Net}}. A high resolution grid is in fact an example of an $\epsilon$-net. There is, however, one key distinction between a grid and general $\epsilon$-nets. Technically, a grid is a covering of the space using $\ell_\infty$ balls, whereas $\epsilon$-nets provide coverings via $\ell_2$ balls. Grids obtain their status as $\epsilon$-nets through norm equivalence, i.e. $\norm{x}_\infty \leq \norm{x}_2 \leq \sqrt{d} \norm{x}_\infty \forall x \in \mathbb{R}^d$. The $\ell_2$ norm is more prevalent than $\ell_\infty$ in the motion planning literature as $\delta$-clearance and connection radius conditions are both stated with respect to Euclidean distance. It stands to reason that general $\epsilon$-nets would provide a more efficient sampling strategy for motion planning problems and algorithms that are specified by $\ell_2$ distance. 
Corollary~\ref{cor:detsuffgrid} providing a worse (i.e. larger) sufficient condition than Theorem~\ref{thm:detsuff} corroborates this intution. Furthermore, in this section we show that, under the condition $\epsilon < \sqrt{\frac{\pi e}{8}} - 1 \approx 0.033$, there exist $\epsilon$-nets that are asymptotically (as $\epsilon \rightarrow 0, d \rightarrow \infty$) more efficient than grids \ifarXiv at \fi covering the space with $\ell_2$ balls. 

\begin{lemma}[Net size and Grid size]\label{lem:netvsgrid}
$\X_{\text{grid}}(w)$ is an $\epsilon$-net if and only if $w \leq \frac{2 \epsilon}{\sqrt{d}}$, thus the smallest grid that is also an $\epsilon$-net is $\X_{\text{grid}} \bigpar{\frac{2\epsilon}{\sqrt{d}}}$. If $\X$ is an $\epsilon$-net obtained from $\hyperref[alg:epsnet]{\texttt{Build-Net}}([0,1]^d, \epsilon)$, then we have
{\small\begin{align*}
\frac{\abs{\X}}{\abs{\X_{\text{grid}}\bigpar{ \frac{2\epsilon}{\sqrt{d}} }}} \leq \sqrt{\pi d} \bigpar{\frac{\sqrt{8} \bigpar{1+\epsilon} }{\sqrt{\pi e}} }^d.
\end{align*}}
\end{lemma}

\ifarXiv
See Section~\ref{pf:lem:netvsgrid} for a proof of Lemma~\ref{lem:netvsgrid}. 
\fi
When $\epsilon < \sqrt{\frac{\pi e}{8}}-1$ we have $\frac{\sqrt{8} \bigpar{1+\epsilon}}{\sqrt{\pi e}} < 1$, hence the ratio given by Lemma~\ref{lem:netvsgrid} goes to zero as $d \rightarrow \infty$. We can then conclude that there exist $\epsilon$-nets that are more efficient than grids for high dimensional problems. For lower dimensions, the upper bound from Lemma~\ref{lem:netvsgrid} is larger than $1$. We suspect however, that the bound is loose in this regime and that $\epsilon$-nets remain competitive with grids even for small dimensions. To test our hypothesis, we conduct an empirical comparison between $\epsilon$-net inspired sample sets and grids in Section~\ref{sec:experiments}.

For some intuition as to why $\epsilon$-nets provide more efficient coverage than grids in high dimensions, we note that Algorithm~\ref{alg:epsnet} returns an $\epsilon$-net $\X$ whose points are pairwise at distance at least $\epsilon$. The grid on the other hand, has spacing at most $\frac{2\epsilon}{\sqrt{d}}$. The shrinkage factor of $O\bigpar{\frac{1}{\sqrt{d}}}$ is precisely from the error in approximating $\ell_2$ by  $\ell_\infty$. Since the minimum separation in the grid is smaller, the balls in $\cup_{x \in \X_{\text{grid}}(2\epsilon/\sqrt{d})} B_2(x,\epsilon)$ overlap more than the balls in $\cup_{x \in \X} B_2(x,\epsilon)$ and thus lead to less efficient coverage. 

\section{Proof Sketches}\label{sec:algorithms}

In this section we sketch the proofs of the sample size conditions in Theorems~\ref{thm:detness} and~\ref{thm:detsuff}. To set the stage, we first discuss the cardinality of minimal $\epsilon$-nets, which play a key role in the proofs. The following result provides upper and lower bounds on the size of minimal nets.

\begin{theorem}[Cardinality of $\epsilon$-nets]\label{thm:netsize}
Let $\A \subset \mathbb{R}^d$ be a set.  
\begin{enumerate}
	\item Every $\epsilon$-net of $\A$ must have cardinality at least $\vol(\A) \sqrt{\pi d} \bigpar{ \sqrt{\frac{d}{2\pi e}} \frac{1}{\epsilon} }^d$.
	\item There exists a $\epsilon$-net of $\A$ with cardinality at most $ \vol \bigpar{\A \oplus B_2 \bigpar{ y, \frac{\epsilon}{2} }} \sqrt{\pi d} \bigpar{ \sqrt{\frac{2d}{\pi e}} \frac{1}{\epsilon}}^d$. 
\end{enumerate}
\end{theorem}
Where $\A \oplus \B := \bigbrace{a+b : a \in \A, b \in \B}$ denotes the Minkowski sum of the sets $\A$ and $\B$. 
\ifarXiv
Theorem~\ref{thm:netsize} is well established result in the learning theory and statistics communities, but for completeness we provide a proof in Section~\ref{pf:thm:netsize}. 
\fi
The $\hyperref[alg:epsnet]{\texttt{Build-Net}}$ procedure outlined by Algorithm~\ref{alg:epsnet} is one particular way to generate an $\epsilon$-net with property $2$ of Theorem~\ref{thm:netsize}.

\subsection{Proof Sketch for Theorem~\ref{thm:detness} }\label{pf:tutorial}
The outline for the proof sketch is as follows. (1) Suppose the sample set $\bigbrace{x_i}_{i=1}^n$ has size $n < \sqrt{\pi d} ( \sqrt{\frac{d}{2\pi e}} \frac{1 - 4\delta}{2\delta})^d$. (2) By part 1 of Theorem~\ref{thm:netsize}, $\bigbrace{x_i}_{i=1}^n$ cannot be a $2\delta$-net of $[2\delta,1-2\delta]^d$. (3) This means that there exists some $y$ so that $\bigbrace{x_i}_{i=1}^n \cap B_2(y,2\delta) = \emptyset$. Furthermore, since $y \in [2\delta,1-2\delta]^d$, we have $B_2(y,2\delta) \subset [0,1]^d$. (4) Consider the motion planning problem $\M := (\F, x_{\text{start}},x_{\text{goal}})$: 
\begin{wrapfigure}{r}{0.2\textwidth}
\ifarXiv
	\includegraphics[width=0.2\textwidth]{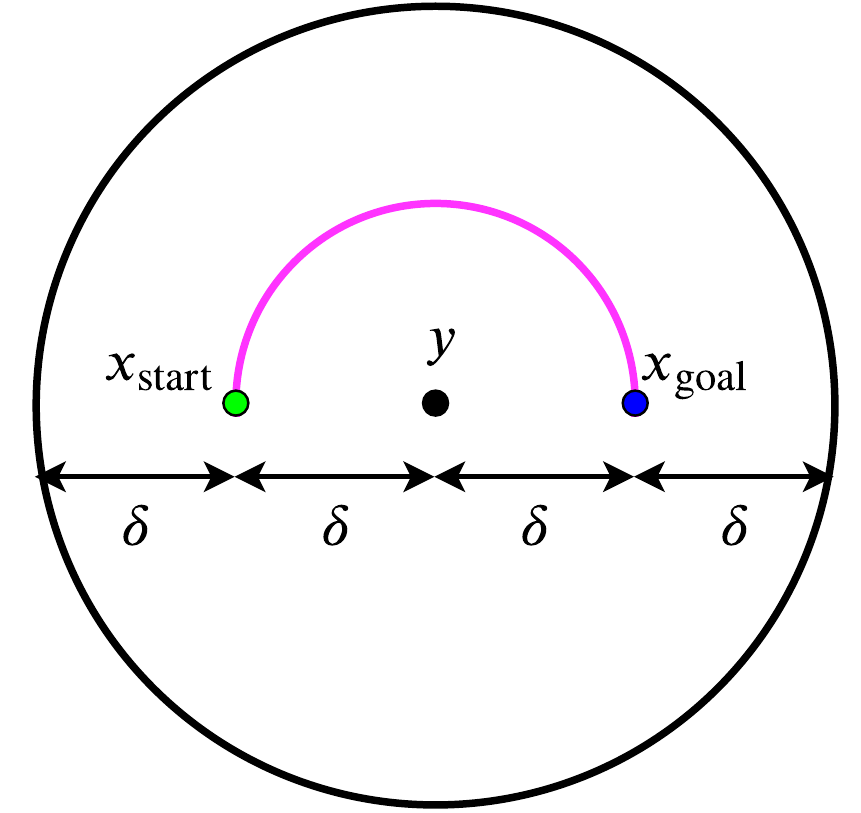}
\fi
\end{wrapfigure}
The black dot is $y$, $x_{\text{start}} = y - \delta e_1, x_{\text{goal}} = y + \delta e_1$ \textit{($\bigbrace{e_i}_{i=1}^d$ are the standard basis vectors for $\dR^d$.)} are depicted by the green and blue dots respectively, and the obstacle set $\overline{\mathcal{F}} := \bigbrace{y} \cup \bigbrace{x : \norm{x-y}_2 = 2\delta}$ is shown in black. This problem is $\delta$-clear, and a solution trajectory is shown in pink. There is, however, no way to reach $x_{\text{goal}}$ from $x_{\text{start}}$ in $G_\M$. This is because the obstacle $\bigbrace{y}$ blocks the edge between $x_{\text{start}}$ and the $x_{\text{goal}}$, and the shell $\bigbrace{x : \norm{x-y}_2 = 2\delta}$ blocks any edges from $\bigbrace{x_{\text{start}},x_{\text{goal}}}$ to $\bigbrace{x_i}_{i=1}^n$, since the points $\bigbrace{x_i}_{i=1}^n$ are outside of $B_2(y,2\delta)$. Thus, the nodes $x_{\text{start}},x_{\text{goal}}$ will have no neighbors in $G_\M$. We conclude that, regardless of the value of $r$, no sample set of size $n < \sqrt{\pi d} ( \sqrt{\frac{d}{2\pi e}} \frac{1 - 4\delta}{2\delta})^d$ can be $(\delta,\infty)$-complete in $[0,1]^d$. The proof of Theorem~\ref{thm:detness} uses this idea but packs the space using a torus instead of a ball. Since a similar argument can be done using the torus, and the torus used has smaller volume than the ball, more tori can be packed, leading to a larger lower bound. 

\subsection{Proof Sketch for Theorem~\ref{thm:detsuff} }\label{sec:sketches:t2}
\begin{figure}
\centerline{
	\ifarXiv
		\includegraphics[scale=0.5]{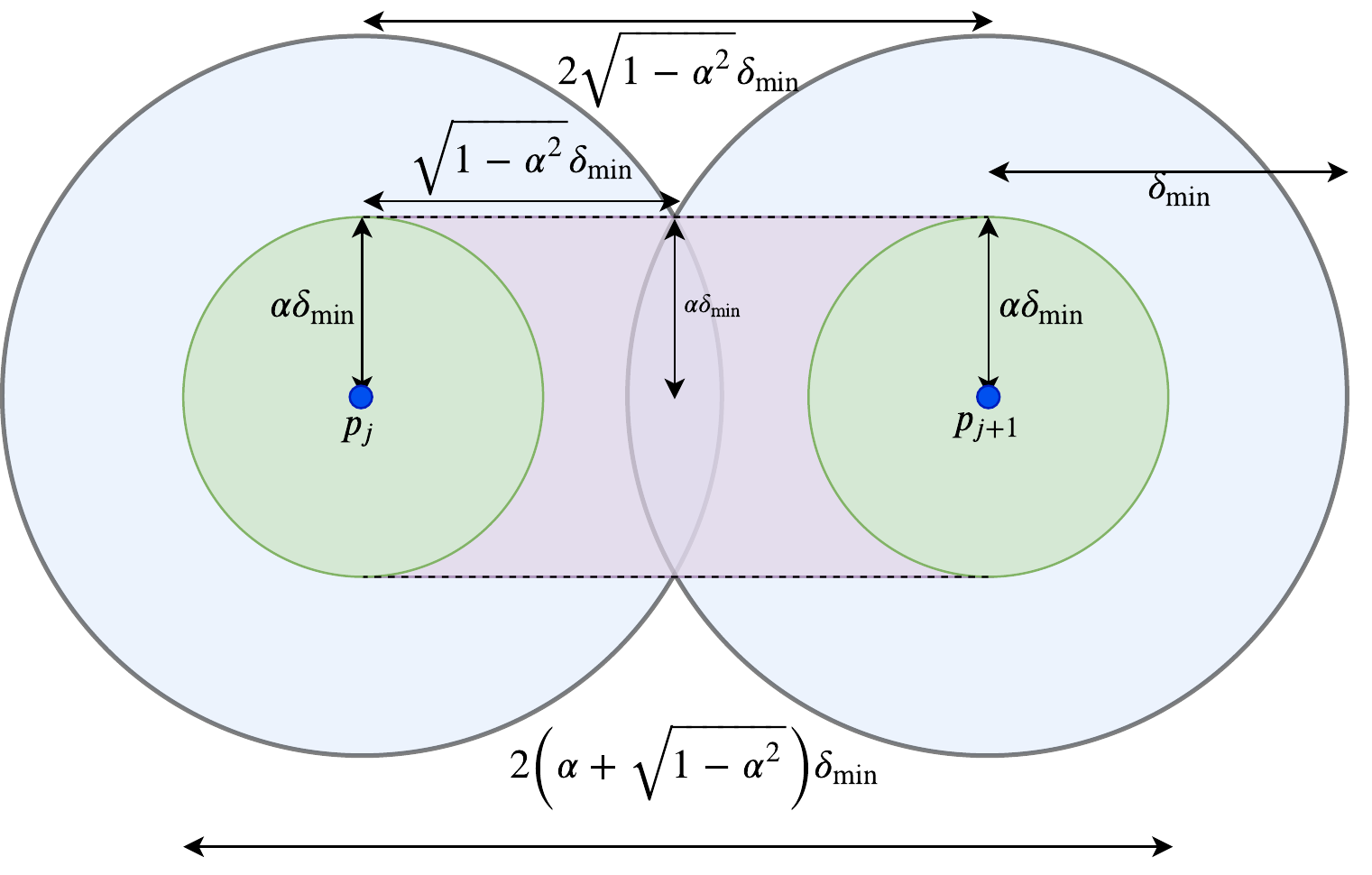}
	\fi
}
\caption{A visualization of the proof technique of Theorem~\ref{thm:detsuff} in 2 dimensions. }\label{fig:detsuff_colorfig_sketch} \vspace{-15pt}
\end{figure}

fSuppose $\M$ is a $\delta_{\text{min}}$-clearance motion planning problem for some $\delta_{\text{min}} > 0$. It then has a shortest $\delta_{\text{min}}$-clear solution path $p$. Consider points $\bigbrace{p_j}_{j=1}^m$ on the path $p$ so that consecutive points are at $2\sqrt{1-\alpha^2} \delta_{\text{min}}$ euclidean distance apart. See Fig.~\ref{fig:detsuff_colorfig_sketch} for an illustration. For each $1 \leq j \leq m-1$, by $\delta_{\text{min}}$-clearance the balls of radius $\delta_{\text{min}}$ centered at $p_j, p_{j+1}$ are collision free. This is illustrated by the blue region in Fig.~\ref{fig:detsuff_colorfig_sketch}. If we have a collection of samples $\X$ which forms a $\alpha \delta_{\text{min}}$-net of the space, then we are guaranteed that there are $z_j, z_{j+1} \in \X$ that are within $\alpha \delta_{\text{min}}$ distance of $p_j$ and $p_{j+1}$ respectively. This is illustrated by the green regions. The condition on $n$ given in the statement of Theorem~\ref{thm:detsuff} in conjunction with Theorem~\ref{thm:netsize} guarantees the existence of such a $\X$. The spacing $2\sqrt{1-\alpha^2} \delta_{\min}$ was chosen so that the convex hull of the green set, depicted as the union of the green and voilet regions, is entirely contained in the blue, collision free region. This means that any line segment joining a point in the left green ball to a point in the right green ball is collision-free. Therefore, for each $j$, the line segment joining $z_j$ to $z_{j+1}$ is collision-free for every $1\leq j \leq m-1$. Furthermore, by triangle inequality, the length of these segments can be at most $2 \bigpar{ \alpha + \sqrt{1-\alpha^2} } \delta_{\min}$. Thus, if we choose $r$ to be this value, then the edges $\bigbrace{\bigpar{z_j, z_{j+1}}}_{j=1}^{m-1}$ will all be in the graph $G_\M(\X,r)$. Since $p$ is a solution path, we can choose the first and last samples $p_1 = x_{\text{start}}$ and $p_m =x_{\text{goal}}$. Therefore the path that concatenates these edges will be a collision-free path from the start to the goal in the graph $G_\M(\X,r)$. Finally, to bound the length of this path, note that the length of the path from $p_j$ to $p_{j+1}$ is $\norm{p_j - p_{j+1}}_2 = 2 \sqrt{1-\alpha^2} \delta_{\min}$. Thus,

\vspace{-4pt}
\begin{footnotesize}
\begin{align*}
\frac{\norm{z_j - z_{j+1}}_2}{\norm{p_j - p_{j+1}}_2} \leq \frac{2 \bigpar{\alpha + \sqrt{1-\alpha^2}} \delta_{\min}}{2\sqrt{1-\alpha^2}\delta_{\min}} = 1 + \frac{\alpha}{\sqrt{1-\alpha^2}} = 1+ \epsilon
\end{align*}
\end{footnotesize}where the last equality is due to the definition of $\alpha$. Since each piece of the path defined by $\bigbrace{z_j}_{j=1}^m$ is not more than $1+\epsilon$ times its corresponding piece in $p$, this gives a path whose total length is at most $1+\epsilon$ times the length of $p$.  

\section{Experiments}\label{sec:experiments}

In this section we demonstrate the improved efficiency of sampling methods that are based on $\epsilon$-nets, compared to grids. While Lemma~\ref{lem:netvsgrid} shows that $\epsilon$-nets are provably more efficient than grids asymptotically as $d \rightarrow \infty$, it does not address the comparison in low dimensions when $d$ is small. To make an empirical comparison in the non-asymptotic regime, we numerically construct sample sets based on $\epsilon$-nets and compare their size and coverage quality to grids.

\subsection{A Sampling Procedure via Templating}\label{sec:experiments:sampling}

\begin{figure}
\centerline{
\begin{tabular}{cc}
\includegraphics[scale=0.23]{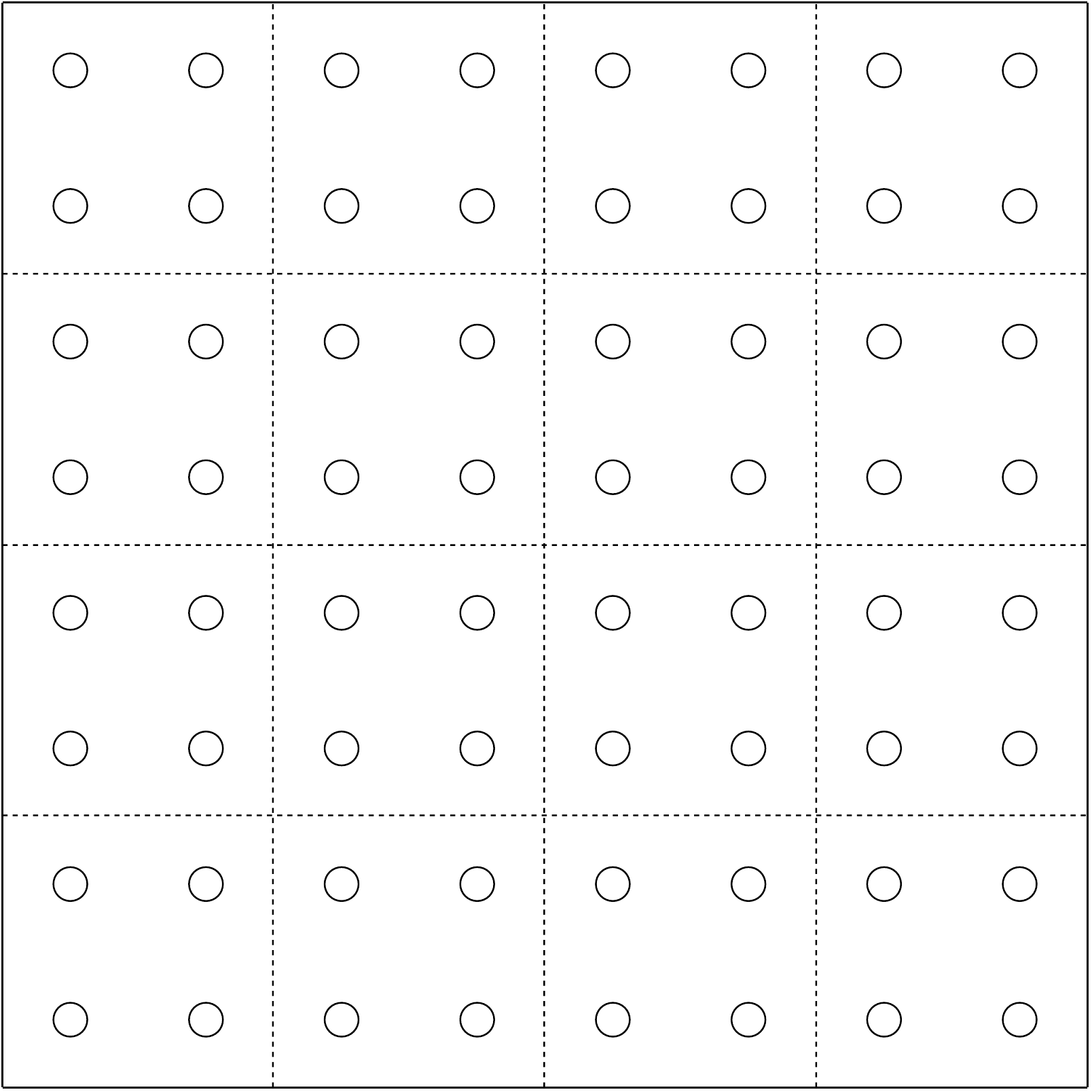} & \includegraphics[scale=0.23]{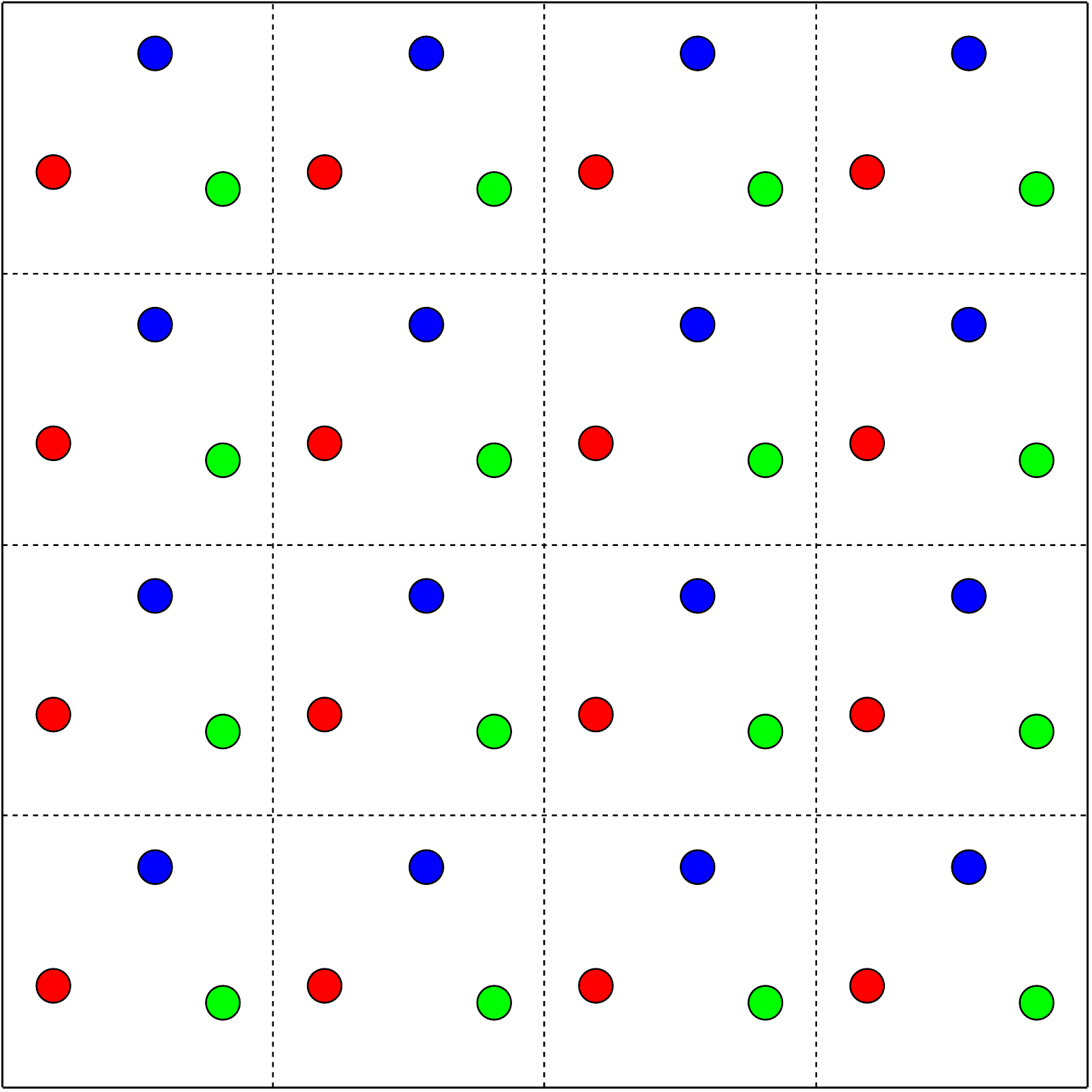}
\end{tabular}
}
\caption{\textbf{Left:} A grid of width $8$ in $d=2$ dimensions can be represented by repeating a grid of width $2$ a total of $16$ times. \textbf{Right:} More generally, any $\T := (\highlight{red}{ x_1}, \highlight{blue}{ x_2}, \highlight{green}{ x_3})$ which forms an $\epsilon$-net of $[0, \frac{1}{4}]^2$ can be replicated 16 times to attain an $\epsilon$-net of $[0,1]^2$. }
\label{fig:grid_fingerprint_2} \vspace{-20pt}
\end{figure}

When $\epsilon$ is small, it is computationally expensive to find a cover of the space with balls of radius $\epsilon$. To alleviate this computational burden, we make the following observation. Suppose $\T$ is a $\epsilon$-net of $[0,m^{-1}]^d$ for some $m\in \dN_+$. Since $[0,1]^d$ is can be written as a union of $m^d$ translations of $[0,m^{-1}]^d$, the union of the analogous $m^d$ translates of $\T$ gives an $\epsilon$-net of $[0,1]^d$. Grids are in fact one example of $\epsilon$-nets with this periodic structure. See Fig.~\ref{fig:grid_fingerprint_2} for an example in $d=2$ dimensions. If instead we want an $\epsilon/2$ net, we can replicate $\frac{1}{2} \T$ a total of $(2m)^d$ times. We call $\T$ a \textit{template}, since scaling it and replicating it appropriately can create $\epsilon$-nets of arbitrary resolution and size. Since rescaling and replication are compuationally cheap, the sample sets can be computed online so long as the templates are precomputed. Due to this versatility, it is acceptable for the computation of these templates to be done offline. 


With this in mind our goal is to find templates of small cardinality for a large range of problem dimensions. Our benchmark will be a grid with spacing $w:=\frac{2\epsilon}{\sqrt{d}}$, since we showed in Lemma~\ref{lem:netvsgrid} that a grid is an $\epsilon$-net if and only if its spacing is at most $w$. The cube $[0,kw]^d$ is covered by a grid of spacing $w$ with exactly $k^d$ points. See Fig.~\ref{fig:grid_fingerprint_2} for a $2$D example. Thus our goal is to find a $\epsilon$-net of $[0,kw]^d$ with fewer than $k^d$ points. This problem is homogeneous in $\epsilon$, so by re-scaling, this is equivalent to finding a $\frac{\sqrt{d}}{2k}$-net of $[0,1]^d$ using fewer than $k^d$ points. Moreover, since grids themselves are periodic, the ratio of sizes of the sample set obtained by repeating $\T$ to the grid will also be $\abs{\T}/k^d$.

Certifying that $\T$ is a $\frac{\sqrt{d}}{2k}$-net of $[0,1]^d$ is nontrivial  because there are infinitely many conditions to satisfy. 
Instead we  sample a large collection of points $\V$ uniformly at random from $[0,1]^d$, and obtain a $\frac{\sqrt{d}}{2k}$-net of those points via the output of \texttt{Build-Net}$(\V, \frac{\sqrt{d}}{2k})$. The resulting set of points $\T$ may not be a $\frac{\sqrt{d}}{2k}$-net of $[0,1]^d$ since it is only certified to be a $\epsilon$-net of the set of densely sampled points. We say a point is uncovered by $\T$ if its nearest neighbor in $\T$ is at distance more than $\frac{\sqrt{d}}{2k}$. 
We verify using Monte Carlo simulation that only a negligible fraction of $[0,1]^d$ is uncovered. 


\subsection{Numerical Results}\label{sec:experiments:results}

We created templates $\T$ using the procedure described in the previous section for various values of dimension $d$ and $k$. We recorded the size of the resulting template and the empirical estimate $\widehat{p}$ on the volume of uncovered space. For each trial, $\widehat{p}$ was computed via Monte Carlo with 10 million samples. The ratio between $\abs{\T}$ and the size of the grid benchmark, $k^d$, is denoted as $\rho$. An efficient sample set $\T$ will lead to a small value of $\rho$. Table~\ref{tbl:template_sizes} summarizes the results of this experiment. 

\begin{table}
\caption{Template sizes for various dimensions}\label{tbl:template_sizes}
\begin{tabular}{|c|ccc|ccc|}
\hline 
& & $k=2$ & & & $k=3$ & \\
\hline
$d$ & $\abs{\T}$ & $\rho$ & $\widehat{p}$ & $\abs{\T}$ & $\rho$ & $\widehat{p}$ \\
\hline
$4$ & $15$ & $0.93$ & $3.7 \cdot 10^{-2}$ & $77$ & $0.95$ & $1.4 \cdot 10^{-2}$ \\
\hline
$5$ & $27$ & $0.84$ & $3.3 \cdot 10^{-2}$ & $189$ & $0.77$ & $5.7 \cdot 10^{-3}$ \\
\hline
$6$ & $57$ & $0.89$ & $5.5 \cdot 10^{-3}$ & $457$ & $0.63$ & $2.4 \cdot 10^{-3}$ \\
\hline
$7$ & $105$ & $0.82$ & $3.1 \cdot 10^{-3}$ & $1078$ & $0.50$ & $1.4 \cdot 10^{-3}$ \\
\hline
$8$ & $173$ & $0.68$ & $1.6 \cdot 10^{-3}$ & $2477$ & $0.38$ & $6.4 \cdot 10^{-4}$ \\
\hline
$9$ & $291$ & $0.58$ & $1.3 \cdot 10^{-3}$ & $5650$ & $0.29$ & $2.6 \cdot 10^{-4}$ \\
\hline
\end{tabular} \vspace{-10pt}
\end{table}

\subsection{Discussion}\label{sec:experiments:discussion}

From Table~\ref{tbl:template_sizes} we see that the relative efficiency compared to the grid, $\rho$, improves as the dimension grows. This observation is consistent with the asymptotic implications of Lemma~\ref{lem:netvsgrid}. For all dimensions, the value of $\widehat{p}$ is on the order of $10^{-3}$, which means that the templates are $\frac{\sqrt{d}}{2k}$-nets of almost the entire space. Notice that the achieved value of $\rho$ is better when $k=3$ than when $k=2$ across all tested dimensions. This is because the \hyperref[alg:epsnet]{\texttt{Build-Net}} procedure does not exploit the fact that $\T$ will be used in a periodic manner. Simply put, there are covering inefficiencies at the boundaries when templates are used in a periodic manner. When covering $[0,1]^d$ with larger templates, there are fewer boundaries between individual template replicas, and thus fewer instances where this boundary inefficiency exists. This reveals a natural trade-off between computation and sample efficiency. Indeed, the cover radius $\frac{\sqrt{d}}{2k}$ is a decreasing function of $k$, meaning it is more difficult computationally to find templates for larger $k$. Thus using small $k$ is computationally cheap, but as Table~\ref{tbl:template_sizes} shows, the quality of the template is worse than what would be obtained for larger $k$. Looking at the extremes, $k=1$ corresponds to the grid, and $k = \frac{\sqrt{d}}{2 \epsilon}$ corresponds to finding an $\epsilon$-net of $[0,1]^d$ without exploiting templates or periodic translation in any way. 

\section{Conclusion and future work}\label{sec:conclusion}

We made progress in the characterization of sample complexity of \prm. We provided lower bounds on the sample size that is necessary for $(\delta,\epsilon)$-complete sampling algorithms. We then complemented the lower bound with achievability results by analyzing $\epsilon$-net and grid based sampling schemes. These sampling schemes are then showed to attain, up to constant factors, the optimal sample complexity for lower dimensional problems. Through numerical experiments we exhibited an $\epsilon$-net inspired sampling strategy, termed templating, that offers nearly the same coverage quality as grids while using significantly fewer samples. 


There are several interesting directions for future research.
First, the gap between the sufficient and necessary conditions for $(\delta,\epsilon)$-completeness is dimension dependent. In high dimensions, the characterization is no longer tight, and the precise dependence of sample complexity on dimension is not yet known. In fact, the gap in our results is due to the gap in characterization of $\epsilon$-net sizes in Theorem~\ref{thm:netsize}, thus closing that gap would have implications for our results. Second, while the templates proposed in our experiments show nearly the same coverage quality as grids, they may not be true $\epsilon$-nets since they are only certified to be an $\epsilon$-net of a large subset of the space. While we showed empirically that the volume of uncovered points is typically on the order of $10^{-3}$, we wish to build templates that are certifiably $\epsilon$-nets for the whole space while retaining reduced size observed in our experiments. Since \texttt{Build-Net} is used in both our theoretical and experimental results, a better algorithm for constructing $\epsilon$-nets would improve results  both in theory and practice. 

\bibliographystyle{IEEEtran}
\bibliography{bibliography} 

\begin{thebibliography}{10}
\providecommand{\url}[1]{#1}
\csname url@samestyle\endcsname
\providecommand{\newblock}{\relax}
\providecommand{\bibinfo}[2]{#2}
\providecommand{\BIBentrySTDinterwordspacing}{\spaceskip=0pt\relax}
\providecommand{\BIBentryALTinterwordstretchfactor}{4}
\providecommand{\BIBentryALTinterwordspacing}{\spaceskip=\fontdimen2\font plus
\BIBentryALTinterwordstretchfactor\fontdimen3\font minus
  \fontdimen4\font\relax}
\providecommand{\BIBforeignlanguage}[2]{{%
\expandafter\ifx\csname l@#1\endcsname\relax
\typeout{** WARNING: IEEEtran.bst: No hyphenation pattern has been}%
\typeout{** loaded for the language `#1'. Using the pattern for}%
\typeout{** the default language instead.}%
\else
\language=\csname l@#1\endcsname
\fi
#2}}
\providecommand{\BIBdecl}{\relax}
\BIBdecl

\bibitem{KavrakiETAL96}
L.~E. Kavraki, P.~Svestka, J.~Latombe, and M.~H. Overmars, ``Probabilistic
  roadmaps for path planning in high-dimensional configuration spaces,''
  \emph{{IEEE} Trans. Robotics and Automation}, vol.~12, no.~4, pp. 566--580,
  1996.

\bibitem{KimmelETAL18}
A.~Kimmel, R.~Shome, Z.~Littlefield, and K.~E. Bekris, ``Fast, anytime motion
  planning for prehensile manipulation in clutter,'' in \emph{18th {IEEE-RAS}
  International Conference on Humanoid Robots, Humanoids 2018, Beijing, China,
  November 6-9, 2018}, 2018, pp. 1--9.

\bibitem{FuETAL19}
M.~Fu, A.~Kuntz, O.~Salzman, and R.~Alterovitz, ``Toward asymptotically-optimal
  inspection planning via efficient near-optimal graph search,'' in
  \emph{Robotics: Science and Systems}, Freiburg im Breisgau, Germany, 2019.

\bibitem{GarrettETAL18b}
C.~R. Garrett, T.~Lozano{-}P{\'{e}}rez, and L.~P. Kaelbling, ``Ffrob:
  Leveraging symbolic planning for efficient task and motion planning,''
  \emph{International Journal of Robotics Research}, vol.~37, no.~1, pp.
  104--136, 2018.

\bibitem{ShomeETAL19}
R.~Shome, K.~Solovey, A.~Dobson, D.~Halperin, and K.~E. Bekris, ``{dRRT*:
  S}calable and informed asymptotically-optimal multi-robot motion planning,''
  \emph{Autonomous Robots}, Jan 2019.

\bibitem{HonigETAL18}
W.~H{\"{o}}nig, J.~A. Preiss, T.~K.~S. Kumar, G.~S. Sukhatme, and N.~Ayanian,
  ``Trajectory planning for quadrotor swarms,'' \emph{{IEEE} Trans. Robotics},
  vol.~34, no.~4, pp. 856--869, 2018.

\bibitem{JSCP15}
L.~Janson, E.~Schmerling, A.~A. Clark, and M.~Pavone, ``Fast marching tree: {A}
  fast marching sampling-based method for optimal motion planning in many
  dimensions,'' \emph{International Journal of Robotics Research}, vol.~34,
  no.~7, pp. 883--921, 2015.

\bibitem{StaETAL15}
J.~A. Starek, J.~V. G{\'{o}}mez, E.~Schmerling, L.~Janson, L.~Moreno, and
  M.~Pavone, ``An asymptotically-optimal sampling-based algorithm for
  {B}i-directional motion planning,'' in \emph{{IEEE/RSJ} International
  Conference on Intelligent Robots and Systems}, 2015, pp. 2072--2078.

\bibitem{SH15}
O.~Salzman and D.~Halperin, ``Asymptotically-optimal motion planning using
  lower bounds on cost,'' in \emph{{IEEE} International Conference on Robotics
  and Automation ({ICRA})}, 2015, pp. 4167--4172.

\bibitem{GSB15}
J.~D. Gammell, S.~S. Srinivasa, and T.~D. Barfoot, ``Batch informed trees
  ({BIT}*): Sampling-based optimal planning via the heuristically guided search
  of implicit random geometric graphs,'' in \emph{{IEEE International
  Conference on Robotics and Automation}}, 2015, pp. 3067--3074.

\bibitem{SolHal16}
K.~Solovey and D.~Halperin, ``Sampling-based bottleneck pathfinding with
  applications to {F}r{\'{e}}chet matching,'' in \emph{European Symposium on
  Algorithms}, 2016, pp. 76:1--76:16.

\bibitem{KavrakiETAL98}
L.~E. Kavraki, M.~N. Kolountzakis, and J.~Latombe, ``Analysis of probabilistic
  roadmaps for path planning,'' \emph{{IEEE} Trans. Robotics and Automation},
  vol.~14, no.~1, pp. 166--171, 1998.

\bibitem{LadKav04b}
A.~M. Ladd and L.~E. Kavraki, ``Fast tree-based exploration of state space for
  robots with dynamics,'' in \emph{Algorithmic Foundations of Robotics}, 2004,
  pp. 297--312.

\bibitem{ChaudhuriKoltun09}
S.~Chaudhuri and V.~Koltun, ``Smoothed analysis of probabilistic roadmaps,''
  \emph{Comput. Geom.}, vol.~42, no.~8, pp. 731--747, 2009.

\bibitem{KF11}
S.~Karaman and E.~Frazzoli, ``Sampling-based algorithms for optimal motion
  planning,'' \emph{International Journal of Robotics Research}, vol.~30,
  no.~7, pp. 846--894, 2011.

\bibitem{SK19}
K.~Solovey and M.~Kleinbort, ``The critical radius in sampling-based motion
  planning,'' \emph{International Journal of Robotics Research}, 2019.

\bibitem{MustafaVaradarajan16}
\BIBentryALTinterwordspacing
N.~H. Mustafa and K.~Varadarajan, ``Epsilon-nets and epsilon-approximations,''
  in \emph{Handbook of Discrete and Computational Geometry}, 3rd~ed., J.~E.
  Goodman, J.~O'Rourke, and C.~D. Toth, Eds.\hskip 1em plus 0.5em minus
  0.4em\relax CRC Press LLC, 2016, ch.~47. [Online]. Available:
  \url{http://www.csun.edu/~ctoth/Handbook/HDCG3.html}
\BIBentrySTDinterwordspacing

\bibitem{LucETAL17}
L.~Janson, B.~Ichter, and M.~Pavone, ``Deterministic sampling-based motion
  planning: Optimality, complexity, and performance,'' \emph{International
  Journal of Robotics Research}, 2017.

\bibitem{TsaoSoloveyPavone20a}
\BIBentryALTinterwordspacing
M.~Tsao, K.~Solovey, and M.~Pavone, ``Sample complexity of probabilistic
  roadmaps via epsilon nets, extended version,'' 2019. [Online]. Available:
  \url{https://drive.google.com/open?id=1Gd6jZwuelcKmUCqt2DV-9G3H9kLlJH-9}
\BIBentrySTDinterwordspacing

\bibitem{SoloveyETAL18}
K.~Solovey, O.~Salzman, and D.~Halperin, ``New perspective on sampling-based
  motion planning via random geometric graphs,'' \emph{International Journal of
  Robotics Research}, vol.~37, no.~10, 2018.

\bibitem{DobsonBekris13}
A.~Dobson and K.~E. Bekris, ``A study on the finite-time near-optimality
  properties of sampling-based motion planners,'' in \emph{{IEEE/RSJ}
  International Conference on Intelligent Robots and Systems, Tokyo, Japan,
  November 3-7, 2013}, 2013, pp. 1236--1241.

\bibitem{DobETAL15}
A.~Dobson, G.~V. Moustakides, and K.~E. Bekris, ``Geometric probability results
  for bounding path quality in sampling-based roadmaps after finite
  computation,'' in \emph{{IEEE International Conference on Robotics and
  Automation}}, 2015, pp. 4180--4186.

\bibitem{SchETAL15}
E.~Schmerling, L.~Janson, and M.~Pavone, ``Optimal sampling-based motion
  planning under differential constraints: The drift case with linear affine
  dynamics,'' in \emph{{IEEE} Conference on Decision and Control}, 2015, pp.
  2574--2581.

\bibitem{SchETAL15b}
------, ``Optimal sampling-based motion planning under differential
  constraints: The driftless case,'' in \emph{{IEEE International Conference on
  Robotics and Automation}}, 2015, pp. 2368--2375.

\bibitem{lavalle06}
S.~M. La{V}alle, \emph{Planning Algorithms}.\hskip 1em plus 0.5em minus
  0.4em\relax Cambridge, U.K.: Cambridge University Press, 2006.

\bibitem{BaschETAL01}
J.~{Basch}, L.~J. {Guibas}, D.~{Hsu}, and {An Thai Nguyen}, ``Disconnection
  proofs for motion planning,'' in \emph{{IEEE International Conference on
  Robotics and Automation}}, vol.~2, 2001, pp. 1765--1772.

\bibitem{McCarthyETAL12}
Z.~{McCarthy}, T.~{Bretl}, and S.~{Hutchinson}, ``Proving path non-existence
  using sampling and alpha shapes,'' in \emph{{IEEE International Conference on
  Robotics and Automation}}, 2012, pp. 2563--2569.

\bibitem{ReidETAL16}
W.~Reid, R.~Fitch, A.~Göktoǧan, and S.~Sukkarieh, ``Motion planning for
  reconfigurable mobile robots using hierarchical fast marching trees,'' in
  \emph{Workshop on the Algorithmic Foundations of Robotics}, 2016.

\bibitem{IchterHarrisonEtAl2018}
B.~Ichter, J.~Harrison, and M.~Pavone, ``Learning sampling distributions for
  robot motion planning,'' in \emph{{IEEE International Conference on Robotics
  and Automation}}, Brisbane, Australia, May 2018.

\bibitem{IchterPavone19}
B.~Ichter and M.~Pavone, ``Robot motion planning in learned latent spaces,''
  \emph{{IEEE} Robotics and Automation Letters}, vol.~4, no.~3, pp. 2407--2414,
  2019.

\bibitem{ChoudhuryBAKRSD18}
S.~Choudhury, M.~Bhardwaj, S.~Arora, A.~Kapoor, G.~Ranade, S.~Scherer, and
  D.~Dey, ``Data-driven planning via imitation learning,'' \emph{International
  Journal of Robotics Research}, vol.~37, no. 13-14, 2018.

\bibitem{Wu2016}
\BIBentryALTinterwordspacing
Y.~Wu, ``Packing, covering, and consequences on minimax risk,'' in \emph{Course
  Lecture notes for ECE598: Information-theoretic methods for high-dimensional
  statistics}, 2016, ch.~14. [Online]. Available:
  \url{http://www.stat.yale.edu/~yw562/teaching/it-stats.pdf}
\BIBentrySTDinterwordspacing

\end{thebibliography}



\ifarXiv
\onecolumn
\section{Proofs}\label{sec:proofs}

\subsection{Proof of Theorem~\ref{thm:detness}}\label{pf:thm:detness}

We first present a simple version of the proof in Section~\ref{pf:tutorial} to help build intution. Section~\ref{pf:thm:detness:sample} presents the proof of the necessary sample size result, which is similar in style to Section~\ref{pf:tutorial} with some more involved technical details. The proof of the necessary connection radius result is given in Section~\ref{pf:thm:detness:radius}.

\subsubsection{Necessary sample size}\label{pf:thm:detness:sample}

\noindent Section~\ref{pf:tutorial} proves a weaker version of Theorem~\ref{thm:detness} by arguing that $(\X,r) \in \texttt{RM}(\delta)$ only if every ball of radius $2\delta$ contains at least one point of $\X$. The proof technique we present here is similar, however, we will be working with rings instead of balls. The resulting necessary condition will be that every ring contained in $[0,1]^d$ must contain at least one point of $\X$. This will lead to a stronger lower bound since the ring we will use is a subset of $B_2(0,2\delta)$, giving rise to a more stringent necessary condition on $\X$ than what was presented in Section~\ref{pf:tutorial}. \\

\noindent To this end, we first define the following functions $y_\delta, y_\delta^\perp : [0,2\pi] \rightarrow \mathbb{R}^d$ entry-wise:
\begin{align*}
\bigpar{ y_\delta(\theta) }_i &= \casewise{
\begin{tabular}{cc}
$\delta \cos (\theta)$ & if $i=1$ \\
$\delta \sin (\theta)$ & if $i=2$ \\
0 & else.
\end{tabular}
}
\text{ and }
\bigpar{y^{\perp}_\delta(\theta)}_i :=
\casewise{
\begin{tabular}{cc}
$-\delta \sin (\theta)$ & if $i=1$ \\
$\delta \cos (\theta)$ & if $i=2$ \\
0 & else.
\end{tabular}
}
\end{align*}
The vectors $\bigbrace{y_\delta (\theta)}_{\theta \in [0,2\pi]}$ form a circle of radius $\delta$ centered at the origin in the subspace spanned by $\bigbrace{e_1,e_2}$, and $y_\delta^\perp(\theta) := \frac{d}{d\theta} y_\delta(\theta)$. Using this definition, we define the ring $\mathcal{R}(x_0, \delta, \delta)$ and the torus $\mathcal{T}(x_0, \delta, \delta)$ as
\begin{align*}
\mathcal{R}(x_0, \delta, \delta) &:= \bigbrace{x_0 + y_\delta(\theta)+ z : \theta \in [0,2\pi], \norm{z}_2 \leq \delta} \\
\mathcal{T}(x_0, \delta, \delta) &:= \bigbrace{x_0 + y_\delta(\theta) + z : \theta \in [0,2\pi], \braket{y_\delta^\perp(\theta), z} = 0, \norm{z}_2 = \delta}.
\end{align*}
The volume of $\mathcal{R}(x_0, \delta, \delta)$ is $(2\pi \delta)(c_{d-1} \delta^{d-1}) = 2\pi c_{d-1} \delta^d$. $\mathcal{T}(x_0, \delta, \delta)$ is the boundary of $\mathcal{R}(x_0, \delta, \delta)$. Our goal is to cover the set $\D$ with rings where 
\begin{align*}
\mathcal{D} := \bigbra{2\delta, 1-2\delta} \times \bigbra{2\delta, 1-2\delta} \times \bigbra{\delta, 1-\delta}^{d-2}.
\end{align*}

\noindent We choose to cover $\D$ because $x_0 \in \D$ is an equivalent condition for $\R(x_0,\delta,\delta) \subset [0,1]^d$. If a set of points $\bigbrace{x_i}_{i=1}^n$ satisfies $\D \subset \bigcup_{i=1}^n \R(x_i, \delta, \delta)$, then
\begin{align*}
\text{vol} \bigpar{\mathcal{D}} &\leq \text{vol} \bigpar{ \bigcup_{i=1}^n \mathcal{R}(x_i, \delta, \delta) } \\
\implies \bigpar{1-4\delta}^2 \bigpar{1-2\delta}^{d-2} &\leq \vol \bigpar{\bigcup_{i=1}^n \mathcal{R}(x_i, \delta, \delta)} \\
&\leq n 2\pi c_{d-1} \delta^d \\
\text{See Section~\ref{note:c_d}} \rightarrow &= n 2\pi \sqrt{\frac{1}{\pi (d-1)}} \bigpar{\sqrt{\frac{2\pi e}{d-1}}}^{d-1} \delta^d \\
\implies n &\geq \sqrt{\frac{2\pi e}{d-1}} \frac{\sqrt{\pi (d-1)}}{2\pi} \bigpar{1 - \frac{2\delta}{1-2\delta}}^2 \bigpar{ \sqrt{\frac{d-1}{2\pi e}} \frac{\bigpar{1-2\delta}}{\delta}}^d \\
&= \sqrt{\frac{e}{2}} \bigpar{1 - \frac{2\delta}{1-2\delta}}^2 \bigpar{ \sqrt{\frac{d-1}{2\pi e}} \frac{\bigpar{1-2\delta}}{\delta}}^d 
\end{align*}
Therefore, if $n < \sqrt{\frac{e}{2}} \bigpar{1 - \frac{2\delta}{1-2\delta}}^2 \bigpar{ \sqrt{\frac{d-1}{2\pi e}} \frac{\bigpar{1-2\delta}}{\delta}}^d$, then for any collection of points $\bigbrace{x_i}_{i=1}^n$, there is a point $x^* \in \mathcal{D}$ so that $x^* \not\in \mathcal{R}(x_i, \delta, \delta)$ for every $1 \leq i \leq n$. This means that $x_i \not \in \mathcal{R}(x^*, \delta, \delta)$ for every $i$. We can show this by contradiction. Suppose $x_i \in \mathcal{R}(x^*, \delta, \delta)$. Then we can write $x_i = x^* + y_\delta(\theta) + z$ for some $\theta \in [0,2\pi]$ and $\norm{z}_2 \leq \delta$. But then choosing $\theta' = \theta + \pi \mod 2\pi$ and $z' = -z$ gives $y_\delta(\theta') = -y_\delta(\theta)$ and we can then write $x^* = x_i + y' + z'$, meaning $x^* \in \mathcal{R}(x_i, \delta, \delta)$ which is a contradiction. \\

\noindent Given this point $x^*$, consider the motion-planning problem 
\begin{align*}
\overline{\F} &= \mathcal{T}(x^*, \delta, \delta) \\
x_{\text{start}} &= x^* + \delta e_1 \\
x_{\text{goal}} &= x^* - \delta e_1.
\end{align*}
The path
\begin{align*}
p(t) := x^* + y_\delta (\pi t), t \in [0,1] 
\end{align*}
is a feasible $\delta$-clearance path, because by definition, $\R(x^*,\delta,\delta)$ is the set of points within $\delta$ of $p$, and obstacles only exist at $\T(x^*, \delta, \delta)$, which is the boundary of $\T(x^*, \delta, \delta)$. This means all points in the obstacle set are at distance at least $\delta$ from all points in $p$. Note that constructing a graph $G$ based on collision free line segments between $\bigbrace{x_i}_{i=1}^n \cup \bigbrace{x_{\text{start}}, x_{\text{goal}}}$ will not have a path from $x_{\text{start}}$ to $x_{\text{goal}}$. This is because $x^* \not \in \F$, (indeed, set $\theta = 0, z = -\delta e_1$), therefore there cannot be an edge directly from $x_{\text{start}}$ to $x_{\text{goal}}$. Thus, a path from $x_{\text{start}}$ to $x_{\text{goal}}$ must include at least one point from $\bigbrace{x_i}_{i=1}^n$. However, by construction, all points in $\bigbrace{x_i}_{i=1}^n$ are in $\overline{\mathcal{R}}(x^*, \delta, \delta)$, while $x_{\text{start}},x_{\text{goal}} \in \mathcal{R}(x^*, \delta, \delta)$. Thus any line segment joining a point in $\bigbrace{x_i}_{i=1}^n$ to $\bigbrace{x_{\text{start}},x_{\text{goal}}}$ must pass through the boundary of the ring, $\mathcal{T}(x^*, \delta,\delta)$, which is an obstacle. Therefore the nodes $x_{\text{start}},x_{\text{goal}}$ are isolated in $G$, i.e. they have no neighbors. Thus there is no path from the start to the goal in $G$ for this obstacle environment. 

\subsubsection{Necessary connection radius}\label{pf:thm:detness:radius}

From Theorem~\ref{thm:netsize}, we know that any $\epsilon$-net of $[\delta, 1-\delta]^d$ must have at least $\sqrt{\pi d} \bigpar{ \sqrt{\frac{d}{2 \pi e}} \frac{1-2\delta}{\epsilon}}^d$ points. Therefore, if $n < \sqrt{\pi d} \bigpar{ \sqrt{\frac{d}{2 \pi e}} \frac{1-2\delta}{\epsilon}}^d$, then there exists no set of $n$ points which is an $\epsilon$-net. Note that
\begin{align*}
n &< \sqrt{\pi d} \bigpar{ \sqrt{\frac{d}{2 \pi e}} \frac{1-2\delta}{\epsilon}}^d \\
\iff \bigpar{ \frac{n}{\sqrt{\pi d}} }^{1/d} &< \sqrt{\frac{d}{2 \pi e}}\frac{1-2\delta}{\epsilon} \\
\iff \epsilon &< (1-2\delta) \sqrt{\frac{d}{2\pi e}} \bigpar{ \frac{\sqrt{\pi d}}{n} }^{1/d}.
\end{align*}
Now for any $r < (1-2\delta) \sqrt{\frac{d}{2\pi e}} \bigpar{ \frac{\sqrt{\pi d}}{n} }^{1/d}$, there cannot exist a $r$-net of size $n$. Therefore, for any $\bigbrace{x_i}_{i=1}^n$, there exists $x^* \in [\delta,1-\delta]^d$ so that $\norm{x_i - x^*}_2 > r$ for every $i \in [n]$. If we choose $x_{\text{start}} = x^*$ and $x_{\text{goal}}$ satisfying $\norm{x_{\text{start}}-x_{\text{goal}}}_2 > r$, then the PRM graph induced by the connection radius $r$ and vertices $\bigbrace{x_i}_{i=1}^n \cup \bigbrace{x_{\text{start}}, x_{\text{goal}}}$ will have no edges connecting to $x_{\text{start}}$, since there are no points within $r$ of $x_{\text{start}}$. As a consequence, the induced graph will not have a path from $x_{\text{start}}$ to $x_{\text{goal}}$. 

\subsection{Proof of Theorem~\ref{thm:detsuff}}\label{pf:thm:detsuff}

In this section we make rigorous the intution presented in Section \ref{sec:sketches:t2}. 

\subsubsection{Proof setup}\label{pf:thm:detsuff:setup}

Fix $\epsilon,\delta \in (0,1)$ and a $\delta$-clear motion planning problem $\M$. Define $\alpha := \frac{\epsilon}{\sqrt{1+\epsilon^2}}$. Suppose we have $n$ satisfying
\begin{align*}
n \geq \sqrt{\pi d} \bigpar{ \sqrt{\frac{2d}{\pi e}} \frac{1-(2-\alpha)\delta}{ \alpha \delta}}^d.
\end{align*}

\noindent Define $\delta_{\text{min}}$ so that 
\begin{align}\label{eqn:tightdelta}
n =  \sqrt{\pi d} \bigpar{ \sqrt{\frac{2d}{\pi e}} \frac{1-(2-\alpha)\delta_{\text{min}}}{\alpha \delta_{\text{min}}} }^d .
\end{align}
Two immediate consequences of this definition are that $\delta_{\text{min}} \leq \delta$ and 
\begin{align}
\delta_{\text{min}} &= \frac{1 - (2-\alpha)\delta_{\text{min}}}{\alpha} \sqrt{\frac{2d}{\pi e}} \bigpar{ \frac{\sqrt{\pi d}}{n} }^{1/d} \nonumber \\
\implies \delta_{\text{min}} &< \frac{1}{\alpha} \sqrt{\frac{2d}{\pi e}} \bigpar{ \frac{\sqrt{\pi d}}{n} }^{1/d}.\label{eqn:radius}
\end{align}

\noindent Let $p :[0,1] \rightarrow [0,1]^d$ be the shortest\footnote{Technically we need to say ``arbitrarily close to the infimum length of a $\delta$-clear path" here.} $\delta$-clear path for $\M$. Choose $\bigbrace{t_j}_{j=1}^m$ so that:
\begin{enumerate}
	\item $t_1 = 0$ and $t_m = 1$. 
	\item $t_j < t_k$ whenever $j < k$.
	\item $\norm{p_j - p_{j+1}}_2 = 2 \sqrt{1-\alpha^2} \delta_{\text{min}}$ for all $1 \leq j \leq m-1$ where $p_j := p(t_j)$. 
\end{enumerate}

\noindent Since $p$ has $\delta$ clearance, $p([0,1]) \subset [\delta,1-\delta]^d$. Since $\delta_{\text{min}} \leq \delta$ we have $[\delta,1-\delta]^d \subset [\delta_{\text{min}},1-\delta_{\text{min}}]^d$. Note that $[\delta_{\text{min}},1-\delta_{\text{min}}]^d \oplus B_2 \bigpar{0, \frac{\alpha \delta_{\text{min}}}{2}}$ is contained in a hypercube of side length $1 - (2-\alpha) \delta_{\text{min}}$. Thus by Theorem~\ref{thm:netsize}, the condition in \eqref{eqn:tightdelta}
ensures that we can choose $\bigbrace{x_i}_{i=1}^n$ to be an $\alpha \delta_{\text{min}}$-net of $[\delta_{\text{min}},1-\delta_{\text{min}}]^d$. In particular, for every $j$, there exists $z_j \in \bigbrace{x_i}_{i=1}^n$ so that $\norm{p_j - z_j}_2 \leq \alpha \delta_{\text{min}}$. \\

\noindent To prove the existence of a path from $x_{\text{start}}$ to $x_{\text{goal}}$ we will show the following two statements are true:

\begin{enumerate}
	\item The line segments connecting $z_j$ to $z_{j+1}$ are collision free for all $1 \leq j \leq m-1$. 
	\item $\norm{z_j - z_{j+1}}_2 \leq 2 \bigpar{\alpha + \sqrt{1-\alpha^2}} \delta_{\text{min}}$. 
\end{enumerate}

\noindent Once we establish these statements, we will choose a connection radius $r$ larger than $2 \bigpar{\alpha + \sqrt{1-\alpha^2}} \delta_{\text{min}}$. This ensures that the edges $\bigbrace{(z_{j},z_{j+1})}_{j=1}^{m-1}$ will all be in the resulting graph $G_\M(\bigbrace{x_i}_{i=1}^n , r)$. The path we construct will then be a concatenate all of the $(z_{j},z_{j+1})$ collision-free edges to get a collision-free path from the start to the goal. 

\subsubsection{Proving $\text{co}\bigpar{\bigbrace{z_j,z_{j+1}}}$ is collision-free}\label{pf:thm:detsuff:sample}
This is equivalent to saying that $\beta z_j + (1-\beta)z_{j+1} \in \F$ for every $\beta \in [0,1]$. Since $p_j, p_{j+1}$ are points on a $\delta$-clear path, the balls $B_2(p_j, \delta), B_2(p_{j+1}, \delta)$ are collision-free. Thus it suffices to show $ \text{co} \bigpar{ \bigbrace{z_j,z_{j+1}} } \subset B_2(p_j, \delta_{\text{min}}) \cup B_2(p_{j+1}, \delta_{\text{min}})$ for each $j$. For any $\beta \in [0,1]$, let $u_\beta := \beta z_j + (1-\beta) z_{j+1}$, and 

\begin{align*}
v_\beta := \underset{v \in \text{co} \bigpar{ \bigbrace{p_i,p_{i+1}} }}{\text{argmin}} \norm{u_\beta -v}_2 .
\end{align*}

By triangle inequality we have
\begin{align*}
\norm{ \beta z_j + (1-\beta)z_{j+1} - \beta p_j + (1-\beta)p_{j+1} }_2 &\leq \beta \norm{z_j - p_j}_2 + (1-\beta) \norm{z_{j+1} - p_{j+1}}_2 \leq \alpha \delta_{\text{min}},
\end{align*} 

and thus $\norm{u_\beta - v_\beta}_2 \leq \alpha \delta_{\text{min}}$. If $v_\beta = p_j$, then $\norm{u_\beta - p_j}_2 \leq \alpha \delta_{\text{min}}$ implies that $u_\beta \in B_2(p_j, \delta_{\text{min}})$, since $\alpha \leq 1$. Similarly, if $v_\beta = p_{j+1}$ then $u_\beta \in B_2(p_{j+1}, \delta_{\text{min}})$. If $v_\beta$ is not one of the endpoints, we can write $v_\beta = \beta^* p_j + (1-\beta^*) p_{j+1}$ where
\begin{align*}
\beta^* &= \arg\min_{\beta \in (0,1)} \norm{ u_\beta - \beta p_j + (1-\beta) p_{j+1} }_2^2.
\end{align*}
As a consequence of the first order optimality conditions, we have $\braket{ p_{j+1} - p_j , u_\beta - v_\beta }=0$. Since the points $\bigbrace{p_j, p_{j+1}, v_\beta}$ are collinear, we also have
\begin{align*}
\braket{ v_\beta - p_j , u_\beta - v_\beta }=\braket{ v_\beta - p_{j+1} , u_\beta - v_\beta }=0.
\end{align*}

\noindent There are now two cases. If $\beta^* \leq 1/2$, we have
\begin{align*}
\norm{u_\beta - p_{j+1}}_2^2 &= \norm{(u_\beta - v_\beta) + (v_\beta - p_{j+1})}_2^2 \\
&= \norm{u_\beta - v_\beta}_2^2 + \norm{v_\beta - p_{j+1}}_2^2 \\
&\leq \alpha^2 \delta_{\text{min}}^2 + \norm{\beta^* p_j + (1-\beta^*) p_{j+1} - p_{j+1}}_2^2 \\
&= \alpha^2 \delta_{\text{min}}^2 + (\beta^*)^2 \norm{p_j - p_{j+1}}_2^2 \\
&= \alpha^2 \delta_{\text{min}}^2 + (\beta^*)^2 4 (1-\alpha^2)\delta_{\text{min}}^2 \\
&\leq \delta_{\text{min}}^2. 
\end{align*}
Otherwise, we have $\beta^* > 1/2$, where we see that
\begin{align*}
\norm{u_\beta - p_{j}}_2^2 &= \norm{(u_\beta - v_\beta) + (v_\beta - p_{j})}_2^2 \\
&= \norm{u_\beta - v_\beta}_2^2 + \norm{v_\beta - p_{j}}_2^2 \\
&\leq \alpha^2 \delta_{\text{min}}^2 + \norm{\beta^* p_j + (1-\beta^*) p_{j+1} - p_{j}}_2^2 \\
&= \alpha^2 \delta_{\text{min}}^2 + (1-\beta^*)^2 \norm{p_j - p_{j+1}}_2^2 \\
&= \alpha^2 \delta_{\text{min}}^2 + (1-\beta^*)^2 4 (1-\alpha^2)\delta_{\text{min}}^2 \\
&\leq \delta_{\text{min}}^2. 
\end{align*}
In either case, $u_\beta \in B_2(p_j, \delta_{\text{min}}) \cup B_2(p_{j+1}, \delta_{\text{min}}) \subset B_2(p_j, \delta) \cup B_2(p_{j+1}, \delta)$. Since this is true for every $\beta \in [0,1]$, the edge joining $z_j$ to $z_{j+1}$ is collision-free. 

\subsubsection{Proving $z_j,z_{j+1}$ are close together}\label{pf:thm:detsuff:radius}

Furthermore, by triangle inquality we have
\begin{align*}
\norm{z_j - z_{j+1}}_2 &= \norm{(z_j-p_j) + (p_j - p_{j+1}) + (p_{j+1} - z_{j+1})}_2 \\
&\leq \norm{z_j -p_j}_2 + \norm{p_j - p_{j+1}}_2 + \norm{p_{j+1} - z_{j+1}}_2 \\
&= \alpha \delta_{\text{min}} + 2\sqrt{1-\alpha^2}  \delta_{\text{min}} + \alpha \delta_{\text{min}} \\
&= 2 \bigpar{ \alpha + \sqrt{1-\alpha^2}} \delta_{\text{min}} \\
&\overset{\hyperref[eqn:radius]{(a)}}{<} \frac{2}{\alpha} \bigpar{ \alpha + \sqrt{1-\alpha^2}} \sqrt{\frac{2d}{\pi e}} \bigpar{ \frac{\sqrt{\pi d}}{n} }^{1/d} \\
&\overset{(b)}{=} 2 \bigpar{1 + \frac{1}{\epsilon}} \sqrt{\frac{2d}{\pi e}} \bigpar{ \frac{\sqrt{\pi d}}{n} }^{1/d}.
\end{align*}

\noindent Where $\hyperref[eqn:radius]{(a)}$ is due to \eqref{eqn:radius} and $(b)$ is due to the definition of $\alpha$. Thus setting $r = 2 \bigpar{ 1 + \frac{1}{\epsilon}} \sqrt{\frac{2d}{\pi e}} \bigpar{ \frac{\sqrt{\pi d}}{n} }^{1/d}$, then the edges $\bigbrace{(z_j,z_{j+1})}_{j=1}^{m-1}$ will all be present in the PRM graph. Concatenating these segments gives a feasible path from $x_{\text{start}}$ to $x_{\text{goal}}$.

\subsubsection{Establishing $\bigpar{1 + \epsilon}$-optimality}\label{pf:thm:detsuff:stretch}
\noindent Finally we bound the length of the path. Denote by $\ell(p)$ the length of $p$, and let $z$ be the path induced by the concatenation of the straight lines connecting $z_1,z_2,\ldots,z_{m-1}$, respectively. 
  \begin{align*}
\ell(z) & =\sum_{j=1}^{m-1} \norm{z_{j+1}-z_{j}}_2 \leq \sum_{j=1}^{m-1} \norm{z_{j+1}-p_{j+1}}_2+\norm{p_{j+1}-p_{j}}_2+\norm{z_{j}-p_{j}}_2 \\ 
& = \sum_{j=1}^{m-1} \norm{p_{j+1}-p_{j}}_2 \bigpar{\frac{\norm{z_{j+1}-p_{j+1}}_2+\norm{p_{j+1}-p_{j}}_2+\norm{z_{j}-p_{j}}_2}{\norm{p_{j+1}-p_{j}}_2}} \\
&\leq \sum_{j=1}^{m-1} \norm{p_{j+1}-p_{j}}_2 \bigpar{1+ \frac{2 \alpha \delta_{\text{min}}}{2 \sqrt{1-\alpha^2}\delta_{\text{min}}}} = \bigpar{1+ \frac{ \alpha}{\sqrt{1-\alpha^2}}} \sum_{j=1}^{m-1} \norm{p_{j+1}-p_{j}}_2 \\
&\leq \bigpar{1+ \frac{ \alpha}{\sqrt{1-\alpha^2}}} \ell(p) \overset{(a)}{=} (1+\epsilon) \ell(p). 
  \end{align*}
where $(a)$ is due to the definition of $\alpha$, which implies that $\frac{\alpha}{\sqrt{1-\alpha^2}} = \epsilon$.

\section{Appendix for Auxiliary Results}

\subsection{Proof of Theorem~\ref{thm:netsize}}\label{pf:thm:netsize}

Suppose $\Y$ is an $\epsilon$-net of $\X$. To show the first part, note that 
\begin{align*}
\X &\subset \bigcup_{y \in \Y} B_2 \bigpar{y, \epsilon} \\
\implies \vol \bigpar{\X} &\leq \vol \bigpar{ \bigcup_{y \in \Y} B_2 \bigpar{y, \epsilon} } \\
&\leq \sum_{y \in \Y} \vol\bigpar{B_2 \bigpar{y, \epsilon}} \\
&= \abs{\Y} c_d \epsilon^d,
\end{align*}
and thus $\abs{\Y} \geq \frac{\vol(\X)}{c_d \epsilon^d}$. \\

\noindent To conclude, we note that $c_d$ is a function of $d$, where the dependence is described in Section~\ref{note:c_d}. Substituting the expression obtained in Section~\ref{note:c_d} into these bounds we obtain

\begin{align*}
\abs{\Y} \geq \vol(\X) \sqrt{\pi d} \bigpar{ \sqrt{\frac{d}{2\pi e}} }^d \frac{1}{\epsilon^d} = \vol(\X) \sqrt{\pi d} \bigpar{ \sqrt{\frac{d}{2\pi e}} \frac{1}{\epsilon} }^d.
\end{align*}

\noindent For the second part, we construct an $\epsilon$-net via Algorithm~\ref{alg:epsnet}. 

For $\Y$ the output of such a procedure, note that by construction, all distinct members of $\Y$ are at distance at least $\epsilon$. Therefore, if $y_1,y_2 \in \Y$ with $y_1 \neq y_2$, we have $B_2(y_1, \epsilon/2) \cap B_2(y_2, \epsilon/2) = \emptyset$. Also note that $y \in \X$ implies that $B_2(y,\epsilon/2) \in \X \oplus B_2(0,\epsilon/2)$. Therefore, we have
\begin{align*}
\X \oplus B_2 \bigpar{ y, \frac{\epsilon}{2} } &\supset \bigcup_{y \in \Y} B_2 \bigpar{ y, \frac{\epsilon}{2} } \\
\implies \vol \bigpar{\X \oplus B_2 \bigpar{ y, \frac{\epsilon}{2} }} &\geq \vol \bigpar{ \bigcup_{y \in \Y} B_2 \bigpar{ y, \frac{\epsilon}{2} } } \\
&= \sum_{y \in \Y} B_2 \bigpar{ y, \frac{\epsilon}{2} } \\
&= \abs{\Y} c_d \bigpar{\frac{\epsilon}{2}}^d,
\end{align*}
therefore $\abs{\Y} \leq \frac{1}{c_d} \bigpar{\frac{2}{\epsilon}}^d \vol \bigpar{\X \oplus B_2 \bigpar{ y, \frac{\epsilon}{2} }}$. Substituting the dependence of $c_d$ on $d$ (See Section~\ref{note:c_d} we obtain 
\begin{align*}
\abs{\Y} \leq \vol \bigpar{\X \oplus B_2 \bigpar{ y, \frac{\epsilon}{2} }} \sqrt{\pi d} \bigpar{ \sqrt{\frac{2d}{\pi e}} \frac{1}{\epsilon}}^d.
\end{align*}

\subsection{Proof of Lemma~\ref{lem:netvsgrid} }\label{pf:lem:netvsgrid}

We have $\abs{\X_{\text{grid}}(w)} = \bigpar{ \frac{1}{w}}^d$, we have
\begin{align*}
[0,1]^d = \bigsqcup_{x \in \X_{\text{grid}}(w)} B_\infty \bigpar{x, \frac{w}{2} },
\end{align*}
therefore $\max_{y \in [0,1]^d} d(y,\X_{\text{grid}}(w)) = \frac{\sqrt{d}w}{2}$. So if we want $\X_{\text{grid}}(w)$ a $\epsilon$-net, then we need $w = \frac{2\epsilon}{\sqrt{d}}$. This gives $\abs{\X_{\text{grid}} \bigpar{ \frac{2\epsilon}{\sqrt{d}} } } = \bigpar{ \frac{\sqrt{d}}{2\epsilon}}^d$. The proof of Theorem~\ref{thm:netsize} gives a construction of a $\epsilon$-net $\X$ of $[0,1]^d$ whose size is $\abs{\X} \leq \text{vol} \bigpar{ [0,1]^d \oplus B_2 \bigpar{0, \frac{\epsilon}{2} } } \sqrt{\pi d} \bigpar{ \sqrt{\frac{2d}{\pi e}} \frac{1}{\epsilon}}^d$. Comparing these two $\epsilon$-nets we see

\begin{align*}
\frac{\abs{\X}}{\abs{\X_{\text{grid}}\bigpar{ \frac{2\epsilon}{\sqrt{d}} }}} &< \bigpar{1+\epsilon}^d \frac{\sqrt{\pi d} \bigpar{ \sqrt{\frac{2d}{\pi e}} \frac{1}{\epsilon}}^d}{\bigpar{ \frac{\sqrt{d}}{2\epsilon}}^d} \\
&= \sqrt{\pi d} \bigpar{ \bigpar{1+\epsilon} \sqrt{\frac{2d}{\pi e}} \frac{1}{\epsilon} \frac{2\epsilon}{\sqrt{d}} }^d \\
&= \sqrt{\pi d} \bigpar{\frac{\sqrt{8} \bigpar{1+\epsilon} }{\sqrt{\pi e}}  }^d 
\end{align*}

\noindent When $\epsilon < \sqrt{\frac{\pi e}{8}}-1$ we have $\frac{\sqrt{8} \bigpar{1+\epsilon}}{\sqrt{\pi e}} < 1$, hence the ratio goes to zero as $d \rightarrow \infty$. We can then conclude that there exist $\epsilon$-nets that are more efficient than grids for high dimensional problems. 

\subsection{Proof for Corollary~\ref{cor:detsuffgrid}}\label{pf:gridsampleruntime}
Throughout this proof we let $\alpha := \frac{\epsilon}{\sqrt{1+\epsilon^2}}$. In the proof of Theorem~\ref{thm:detsuff} from Section~\ref{pf:thm:detsuff}, we showed that if $\X$ is a $\alpha \delta$-net and $r \geq 2 \bigpar{ \alpha + \sqrt{1-\alpha^2} } \delta$, then $(\X,r) \in \texttt{RM}(\delta)$ with $\delta$-stretch at most $1+\epsilon$. Due to norm equivalence, to obtain a $\alpha\delta$ net using a grid, the grid spacing $w$ must be at most $\frac{2\alpha\delta}{\sqrt{d}}$. Since the solution path $p$ is $\delta$-clear, it must be at distance at least $\delta$ from the boundary of $[0,1]^d$, hence it is entirely contained in $[\delta,1-\delta]^d$. Thus to ensure $\delta$-completeness with stretch at most $1+\epsilon$, we need to construct a grid of $[\delta,1-\delta]^d$ with spacing $\frac{2\alpha\delta}{\sqrt{d}}$. A grid of $[\delta,1-\delta]^d$ with spacing $\frac{2\alpha\delta}{\sqrt{d}}$ has exactly $\frac{\sqrt{d}(1-2\delta)}{2\alpha\delta}$ points along each axis, hence 
\begin{align*}
\abs{\X_{\text{grid}} \bigpar{ \frac{2\alpha\delta}{\sqrt{d}} } } = \bigpar{ \frac{\sqrt{d}(1-2\delta)}{2\alpha\delta} }^d.
\end{align*}
\subsection{Unit Ball Volume Approximation}\label{note:c_d}

We begin by defining the gamma function $\Gamma : \mathbb{R}_+ \rightarrow \mathbb{R}_+$ as
\begin{align*}
\Gamma(x) := \int_0^\infty t^{x-1} e^{-t} dt.
\end{align*}
A simple consequence of integration by parts is that for positive integer $n$, $\Gamma(n) = (n-1)!$. For general real numbers, the gamma function has no closed for expression. We do have good approximations, described by the following result:

\begin{theorem}[Stirling's Approximation]\label{thm:stirling}
\begin{align*}
\lim_{x \rightarrow \infty} \frac{\Gamma(x+1)}{\sqrt{2\pi x} \bigpar{ \frac{x}{e} }^x } = 1,
\end{align*}
and we have the following bounds for finite $x$:
\begin{align*}
\sqrt{2\pi x} \bigpar{ \frac{x}{e} }^x \leq \Gamma(x+1) \leq e \sqrt{x} \bigpar{ \frac{x}{e} }^x.
\end{align*}
\end{theorem}
\noindent It is known that the volume of the unit ball in $d$ dimensions is
\begin{align*}
c_d := \frac{\pi^{d/2}}{\Gamma \bigpar{ \frac{d}{2} + 1 } }.
\end{align*}
we obtain the desired result by applying Stirling's approximation to bound $c_d^{-1}$. 
\begin{align*}
c_d^{-1} = \frac{\Gamma \bigpar{ \frac{d}{2} + 1 }}{\pi^{d/2}} \sim \sqrt{2\pi\frac{d}{2}} \bigpar{ \frac{d}{2 e} }^{d/2} \frac{1}{\pi^{d/2}} = \sqrt{\pi d} \bigpar{ \sqrt{\frac{d}{2 \pi e}} }^{d} .
\end{align*}	
\fi

\end{document}